# City representation in the Soviet propaganda: quantifying biases of the Soviet worldview

**Authors**

Mikhail V. Tamm,[1]* Mila Oiva,[2]† Ksenia D. Mukhina[3], Mark Mets[2], Maximilian Schich[3]

**Affiliations**

[1] School of Digital Technologies, Tallinn University, Tallinn, Estonia.

[2] School of Humanities, Tallinn University, Tallinn, Estonia

[3] Baltic Film and Media School, Tallinn University, Tallinn, Estonia

† currently at University of Turku, Turku, Finland

* corresponding author, e-mail thumm.m@gmail.com

**Abstract**

Cultural data typically contains a variety of biases. In particular, geographical locations are unequally portrayed in media, creating a distorted representation of the world. Identifying and measuring such biases is crucial to understand both the data and the socio-cultural processes that have produced them. Here we suggest measuring geographical biases in a large historical news media corpus by studying the representation of cities. Leveraging ideas of quantitative urban science, we develop a mixed quantitative-qualitative procedure, which allows us to get robust quantitative estimates of the biases. These biases can be further qualitatively interpreted resulting in a hermeneutic feedback loop. We apply this procedure to a corpus of Soviet newsreel series 'Novosti Dnya' (News of the Day) and show that city representation grows super-linearly with city size and is further biased by city specialization and geographical location. This allows to systematically identify geographical regions which are explicitly or covertly emphasized by Soviet propaganda and quantify their importance.

## Introduction

Large scale cultural data is subject to quantitative patterns originating from complex interplay of general trends, local specifics and selection biases. These patterns, which reveal the innate stereotypes and biases of the surroundings producing the data, in many cases lend themselves to measurement and interpretation. This is especially clear in the study of the fortune of urban centers in cultural history [1-6].

Representation and selection biases in the cultural data are crucial for understanding such quantitative patterns. One example is unequal representation of geographical locations in the media, which might, among other things, lead to neglect or depreciation of narratives crucially important to some national, geographic or social groups. The ability to pinpoint and quantify this unequal representation can give essential insights into the underlying normative worldview of the media-producing society. Attention to different geographical locations in the media has been studied for a long time [7-10]. Visibility of a country in the international news scene is known to be influenced by multiple reasons[9]: ad-hoc political and economic events and regional centrality can be reasons for over-representation, while under-representation can be driven by peripheral geographical position and by invisible conflicts. Administrative status, economic development, number of central institutions, tourist resources and distance to the capital may affect the amount of online media attention received by cities as shown for contemporary China[11]. Population adjusted Tweet density is known to be lower in ``left behind'' areas[12].

While geographical and spatial biases are present at all spatial scales, from continents to neighborhoods, we argue that cities form a natural probe to study representation of geographical space in historical media. Cities are numerous, their size is relatively well-defined, spans multiple orders of magnitude, and is reasonably well-documented historically[13,14]. Recent influx of ideas from complex systems theory into urban science, especially the idea of urban scaling [4,15-18] (see also books [19-21] and recent reviews [22,23]) provides a useful conceptual framework for understanding the city representation.

Here we provide a general procedure for extracting insights regarding the representation of geographical space in historical media from the data on how cities are mentioned in a historical news corpus. Our method consists of following feedback-loop-forming steps (see Fig. 1): (i) formulation of a hypothesis about parameters governing city representation; (ii) calculation of the parameters of the hypothesis by minimization of an explicitly defined loss function, (iii) elimination of irrelevant parameters based on a predetermined information-theoretic criterion, and (iv) correction of the hypothesis based on qualitative analysis of the outliers.

We exemplify this procedure by the systematic study of the corpus of Soviet newsreels titled "Novosti Dnya" (News of the day)[24]. Newsreels - short news films shown in cinemas before the evening's feature film - were influential means of depicting the world for the cinema goers in the 20th century, visualizing events, individuals and places that the spectators could read about in the newspapers. Throughout almost all history of the Soviet Union, the production system and censorship made sure that newsreels reflected the policies of the leadership. It was an openly acknowledged principle that news production was to serve the goal of building communism by representing the contemporary world accordingly, i.e. the goals of objectivity or impartiality were explicitly rejected, and the purpose of news was to show the events in the light of innate social, economic, political, and cultural superiority of the communist system [25]. Thus, the content of newsreels reflects the prescribed worldview, the set of topics, places and individuals, which were considered appropriate to be presented and discussed in an official source. They provide therefore an interesting glimpse into the history of the Soviet Union and its political and media culture.

The spatial history of Russia and the Soviet Union has identified entanglements of imperial politics, practices, and identities with spatiality [26-29], including the persistent connection of the Russian and Soviet imperial identities with the idea of the vastness of the country, covering "one sixth of the world" [30,31], the connection of imperial visions and territorial expansion [32], Eurasianism and its political interconnections [33], and the spatial and geographical arrangements during the Stalinist era (late 1920s to early 1950s)[34-37]. Despite the official Soviet ideology of equality, interconnected social and spatial hierarchies were at the core of the Soviet system. These hierarchies originated in both the political ideology and the pragmatic considerations of usefulness for the state and were entangled with spatial hierarchy, where Moscow and major cities were at the top, capitals of the Soviet republics at the second tier, and small cities far away from Moscow were at the bottom [34,38,39]. Likewise, the Soviet media system was hierarchically and geographically organized with most important newspapers, radio stations and film studios located in Moscow [40].

Following the general principle of politicization of news, representation of the outside world in the Soviet media was determined by current politics, and its shifting tendencies of isolationism or expansionism [25,41]. The Soviet Union was depicted as a focal point of world history, its socialist allies were seen as "younger brothers", following the lead of the Soviet Union, and the whole socialist camp - as surrounded by capitalist enemies, shaken by social hardships[41]. However, since the mid-1950s the Soviet culture started to open up to the outer world[31,40, 42-43], and the presentation of both socialist "allies" and capitalist "enemies" was further graduated according to how friendly the relations with a particular foreign country were[44].

## Results

We start with counting the number of news stories in the corpus of the Soviet Newsreel ``Novosti Dnya'' (see Methods section for the details on the dataset) mentioning different cities. Table 1 summarizes the results for the seven largest cities inside and outside the USSR plus three most mentioned cities outside the top 7. Notably, mentions of Soviet cities are systematically larger than those outside; correlation between mentions and population is much clearer for Soviet cities. Moscow is a big outlier, partly due to its capital functions, partly because of easier access to locations within it for a Moscow-based newsreel. Given these observations, we exclude it from further analysis (see, however, more detailed discussion in the SM), and consider mentions of Soviet and foreign cities separately.

### *Cities in the USSR*

*Population only model*. To study representation patterns of the Soviet cities, we collected the data on all 309 cities with population above 0.03% of the population of the USSR (this threshold is chosen because it roughly corresponds to 1 mention per city), see Fig. 2. Our first hypothesis, in the spirit of urban scaling theory, is that

$$m_{\mathrm{I},i}(\boldsymbol{K}, P_i) = c\, P_i^{a},\ \log m_{\mathrm{I},i} = \log c + a \log P_i \quad (1)$$

where the parameter $P_i$ is the rescaled average fraction of the USSR population living in this city

$$P_i = \frac{1}{3\times10^{-4}} \frac{1}{3} \left( \frac{Pop_{i,1959}}{Pop_{\mathrm{USSR},1959}} + \frac{Pop_{i,1970}}{Pop_{\mathrm{USSR},1970}} + \frac{Pop_{i,1979}}{Pop_{\mathrm{USSR},1979}} \right) \quad (2)$$

and numerical constants $\boldsymbol{K} = \{c, a\}$ are obtained by maximal likelihood estimation (M3) and equal to $c = 1.34 \pm 0.12$, and $a = 1.33 \pm 0.04$ (here and below we provide 95% confidence intervals). Notably, the scaling constant $a$ is larger than 1, indicating an agglomeration effect [4, 45].

As shown in Figs. 2 and 3A, while the model describes the majority of cities reasonably well, there is a significant number of outliers: 30 (10%) with p-value below 0.001, and 39 (13%) more with p-value between 0.001 and 0.05 (see [46] for the full list). Many of the outliers are geographically clustered, some others share industrial specialization (e.g., hydroelectricity and steelworks). To allow for that, we constructed two competing models, one allowing for city geographical location, another – for city specializations.

*Geolocation model*. The hypothesis here is that the USSR was split into geographical regions with different intrinsic levels of representation, assuming the expectation (M1) to take form

$$\log m_{\mathrm{II},i} = \log c + a \log P_i + \sum_\alpha I_{i,\alpha} \log k_\alpha \quad (3)$$

where $I_{i,\alpha}$ is the indicator function of $i$-th city belonging to $\alpha$-th region, $\sum_\alpha I_{i,\alpha} = 1$.

*A priori* the set of relevant geographic regions is unknown. To determine it, we start with a set of 40 seed regions (see Fig. 4A), including all Union-level republics separately. In order to avoid too small groupings, we add top 5 cities of each Union-level republics to the list, which increases the total number of cities of interest to 328. Three larger republics, Kazakhstan, Russia and Ukraine, are further split into subregions.

We then apply the parameter-removing procedure outlined in the Methods section: starting with a given set of regions, we choose a region with the smallest number of cities in it, attempt to merge it with each of the geographically adjacent regions, and check if the condition (M6) is satisfied. If it is not, the merger with the smallest decrease of the loss function is accepted. The procedure is continued until no further merges are possible. Fig. 4B shows the resulting set of regions and corresponding ranges of $k_\alpha$ (see [46] for details).

Most over-represented regions are the vicinity of Moscow (probably due to convenience of filming there), Baltic States, South (including Southern Russia, Georgia and Crimea), North-East and Northern Kazakhstan. The over-representation of North-East may be related to its attractiveness as a faraway exotic place and to a large per capita concentration of the ideologically important "stroyki kommunizma" (construction projects of communism) – big development projects often located in the East: note that Bratsk and Krasnoyarsk – location of two important "stroyki kommunizma" in Eastern Siberia – are also overrepresented. Northern Kazakhstan is the region where "osvoyenie tseliny" (reclamation of virginlands), the major political campaign of 1950s, took place. The central role this campaign played in the biography of L. Brezhnev ensured North Kazakhstan remained important for the official mythology of the later Brezhnev era. Over-representation of South and Baltics might be related to the cultural attraction of the 'Soviet West' [47, 48] or to recreational attractiveness of this regions for the film crews.

The most under-represented regions are Western Urals, Western Siberia and Russia-Ukrainian borderlands (Donbas in Ukraine and Rostov oblast in Russia). All three are industrial heartlands lacking a clear ideological significance beyond their industrial role.

*Specialization model*. An alternative approach to understanding city representation is to study how it is correlated with the presence of some industries or administrative functions. The hypothesis in this case is that expectation (M1) takes the form

$$\log m_{\mathrm{III},i} = \log c + a \log P_i + \sum_\beta I_{i,\beta} \log s_\beta \qquad (4)$$

where index $\beta$ enumerates specializations, $I_{i,\beta}$ equals 1 if specialization $\beta$ is present in the *i*-the city and 0 otherwise (cities can have more than 1 specialization, or no specialization at all). We start with 19 hypothetical specializations, and reduced their number, either by elimination or by merging them together (see[46]) according to the same information-theoretic rule as above. Seven out of 19 specializations turned out to be relevant, see Fig. 4C and Table 2. Naturally, the administrative and symbolic value of a capital of a Union-level republic results in higher representation. On the contrary, the capitals of nation-based administrative-units inside the Russia proper are under-represented both with respect to centers of non-national administrative units and to cities with no administrative function. There is a significant boost for seaside cities but only for the Baltic and Black seas and the Pacific, presumably due to their strategic, cultural and/or recreational importance. Strikingly, while steelworks and huge hydroelectric dams boost representation, there is no similar effect for, e.g., automobile industry.

*Full model.* The geolocation and specialization models give two different angles for interpretation of the representation of Soviet cities, each of them can explain some of the regularities in data, but not others. Moreover, there are correlations between them: capitals of the republics are concentrated along the western and southern borders, there are plenty of seaports in the Baltic and South regions, many steelworks in the Eastern Urals, etc. To study the interplay of specialization and location we introduce a combined model:

$$\log m_{\mathrm{IV},i} = \log c + a \log P_i + \sum_{\alpha} I_{i,\alpha} \log k_{\alpha} + \sum_{\beta} I_{i,\beta} \log s_{\beta} \quad (5)$$

where index $\alpha$ enumerates geographical regions (with same seed regions and merging procedure) and $\beta$ enumerates specializations (only 7 specializations which proved relevant are used). The results are summarized in Fig. 4D and in Table 2. Interplay between geographical and specialization parameters notably leads to a smaller number of relevant geographical regions.

While the Moscow region and Northern Kazakhstan retain their prominence, the significant over-representation of the Baltic region in the geolocation model is fully explained by the effects of the Baltic Sea and the capital status of Riga, Tallinn, and Vilnius. Other

Western regions of the USSR (Belarus, Moldova and Western Ukraine) are also mentioned similarly to the generic parts of Russia proper, with over-representation of Minsk and Chisinau explained by their capital status.

In turn, "oriental" republics of Central Asia and Transcaucasus are even more significantly under-represented after control for the republic capitals, emphasizing the Eurocentric nature of the Soviet ideological system. Also underrepresented is the part of Russia interjacent between the European Center and the ideologically important East. Similarly, under-representation of Central and Eastern Ukraine and the Rostov region of Russia may point to an ambiguous intermediate status of Ukraine in the implicit Soviet nomenclature of ethnicities.

In Fig. 3B predictions of the full model are compared with actual mention for individual cities. Naturally, the results are still scattered around the predicted values but with a much narrower spread than in Fig. 3A. There are just 10 (3%) cities with $p < 0.001$ and 29 (9%) with $0.05 > p > 0.001$, compared with 10% and 13%, respectively, for the population only model and, on the other hand, with 0.1% and 5%, respectively, expected if the formulae for the expected values were exact.

### *Cities outside the USSR*

The newsreel "Novosti Dnya" mostly specialized in the internal Soviet news. As a result, foreign cities were mentioned more rarely than Soviet ones (see Table 1). However, it is possible (see [46] for full details) to construct a city-representation model in a methodologically similar way. Given the sparseness of the dataset the only relevant city specialization seems to be capital status, which is to be expected given that country-related political news are often coming from the capital. The optimal formula for the expected number of mentions of the foreign cities is

$$\log m_{\mathrm{F},i} = \log c + a\left(\log P_i + \frac{1}{2} I_{i,cap} \log \frac{P_{i,c}}{P_i}\right) + \sum_\alpha I_{i,\alpha} \log k_\alpha \quad (8)$$

where $P_i$ is the population of the $i$-th city, $P_{i,c}$ is the population of the country to which it belongs, $I_{i,cap}$ is the indicator function of a city being a capital, $I_{i,\alpha}$ is the indicator function of belonging to a geographical group $\alpha$, while residual $c$, scaling exponent $a$, and boost factors $k_\alpha$ are numerical constants to be determined by the maximal likelihood estimate. The number and composition of the geographical groups is optimized similarly to the Soviet cities case. The treatment of the capital status by replacing the city population with the geometric mean of city and country populations is a result of optimization of a more complex formula[46].

The list of relevant geographical groups, as well as optimal values of $c, a, k_\alpha$ are given in Table 3. The inequality between geographical groups is much stronger here than for the cities inside the USSR. There is a clear difference in representation of three groups of countries, easily identifiable with the classification into “first” (developed capitalist: Europe, USA, Canada and Australia), “second” (socialist) and “third” (developing) worlds commonly used in the contemporary Soviet sources. Interestingly, the optimization algorithm confidently classifies Japan and China in the latter group. This might signify once again the intrinsic Eurocentric (or even white-centric) bias in the worldview of newsreel creators. Both developed capitalist and socialist camps split further into three groups with significantly different levels of representation.

The tiers in the socialist world can be explained by a combination of the level of ideological conformity of corresponding regimes and the Eurocentrism of the Soviet worldview. Indeed, lower tier consists of three non-European socialist countries plus

Yugoslavia and Romania, which had strained relations with the Soviet Union, while the top tier includes Mongolia, Bulgaria and Czechoslovakia, whose authorities toed the Soviet line exceptionally closely, except for several months of the Prague Spring in the case of Czechoslovakia.

The tiers in the capitalist world point once again to the Eurocentrism, and to the extreme importance of neutral European countries, Finland and Austria, as the Soviet "windows to the West"[49-51].

**Discussion**

In this paper we analyze the representation of geographical space in historical Soviet propaganda media using a predominantly post-Stalin era corpus of the "Novosti Dnya" newsreel series. Our analysis is based on quantitative models of city mentioning and allows to elucidate and quantify biases in city representation. Full interpretation of these biases needs further qualitative analyses of the corpus, coupled with other topical historical sources. However, we observe the following important repeating motives.

Our corpus shows a clear hierarchy of representation with the Soviet Union on top, followed by the Socialist block, the developed capitalist world and, finally, the developing world. Representation of cities grows superlinearly with city population, indicating positive agglomeration effects, and is boosted by capital status (national or Soviet-Union-level republican).

Contrary to the messaging of the official Soviet ideology, which emphasized equality of nations and anticolonial movement, the silently sold Soviet worldview is heavily centered on Europe being in the role of a privileged or hierarchically higher "Other"[52]. In agreement with previous qualitative observations[34, 39, 44, 48], we find that European countries (both socialist and capitalist) are mentioned more than their counterparts elsewhere, with the only exception of Mongolia. The same dynamics holds within the USSR: western republics (with a profound exception of Eastern and Central Ukraine) are mentioned much more than those in Central Asia and Southern Caucasus. We found that this profound East-West asymmetry is surprisingly under-reported in the post-colonial studies of the USSR.

Some regions and branches of heavy industry have an outsized ideological importance. Regional examples are Northern Kazakhstan inside the USSR, the most loyal countries of the socialist block, and, most strikingly, the two neutral capitalist countries, Austria and Finland.

Seemingly, Soviet worldview deliberately avoids mixed and intermediate cases and situations: while a trait is celebrated and emphasized in its fully developed form (huge dams, faraway North-East location, Union-level capital status), intermediate forms of the same trait (medium-sized dams, location in West Siberia or Urals, capital of lower-level national autonomy) are often under-represented. It is possible that some similar mechanism is behind the under-representation of Eastern and Central Ukraine.

Finally, in some cases places are overmentioned seemingly just because it is convenient (close to Moscow) or pleasant (Baltic and Black Sea coasts) to film there.

While studying a particular example of a Soviet media corpus, we develop a general approach to extracting information on geographical biases from historical news corpora. The suggested procedure combines quantitative and qualitative steps into a single feedback loop, allowing to systematically refine hypotheses about relevant factors and to measure biases in robust quantitative way. The methodology developed here can be used for the analysis of multiple other datasets, including historical newspaper and more recent online media corpora, and hopefully will become a standard in the field. Speaking more generally, we show here how combination of relatively simple reference models routed in the complexity theory and rigorous statistical analysis of deviations from those models can be leveraged to extract significant new information in such traditionally qualitative fields as history and media studies.

# Methods

### Hypothesis and loss function

The method we develop here aims to extract the quantitative estimates of the factors determining the frequency of mentions of the cities in a robust and reliable way. Input consists of a list of $N$ cities with the numbers of times $n_i$ $(i = 1, \ldots, N)$ they are mentioned, and a *hypothesis*, i.e. an assumption that the expected number of mentions

$$m_i = m(\boldsymbol{A}_i, \boldsymbol{K}) \quad \text{(M1)}$$

is a certain function of a vector of city attributes $\boldsymbol{A}_i = \{A_{i,1}, \ldots, A_{i,m}\}$ (e.g., city population or binary variables like whether the city serves a certain administrative function) and a vector of numerical parameters $\boldsymbol{K} = \{K_1, \ldots, K_l\}$, common for all cities. We assume that actual number of mentions $n_i$ is a Poisson random variable with mean $m_i$, and that mentions of different cities are independent. This implies the loss function

$$L(\{n_i\}, \{\boldsymbol{A_i}\}, \boldsymbol{K}) = \sum_{i=1}^{n} \log p\big(n_i, m(\boldsymbol{A}_i, \boldsymbol{K})\big), \; p(n, m) = 2 \min\left[\frac{\Gamma(n+1,m)}{n!}, 1 - \frac{\Gamma(n,m)}{(n-1)!}\right], \quad \text{(M2)}$$

where $\Gamma(n, m)$ is the incomplete gamma-function, so that the cumulative distribution function of a Poisson distribution with average *m* is $\Gamma(n, m)/(n-1)!$. Function $p(n, m)$ in (2) estimates how improbable it is to observe a value *n* of a variable, whose expectation is *m*, i.e. it is the p-value for the Poisson distribution.

For a given hypothesis $m(\boldsymbol{A}, \boldsymbol{K})$ the optimal values of the parameters $K_0$ come from the maximal likelihood estimate

$$K_0 = \mathrm{argmax}_{\boldsymbol{K}} L(\{n_i\}, \{\boldsymbol{A}_i\}, \boldsymbol{K}). \quad \text{(M3)}$$

Contrary to the commonly used least-square method, this procedure, inspired by [53], explicitly accounts for the difference in the scale of fluctuations for frequently and rarely mentioned cities, including cities with zero mentions[54].

### Confidence intervals

To estimate the single-parameter confidence intervals we assume that probability $\Pi(k)$ of observing a given value $k$ of a parameter $K_i \in \boldsymbol{K}$ is proportional to

$$\Pi(k) \sim \exp \Lambda(k)\,, \quad \Lambda(k) = \max_{\boldsymbol{K'}} L(\{n_i\},\{\boldsymbol{A}_i\},k,\boldsymbol{K}') \qquad \text{(M4)}$$

where $\boldsymbol{K}'$ is a set of all parameters in $\boldsymbol{K}$ except $K_i$ , and we assume that in the vicinity of its global maximum $k_0$ $\Lambda(k)$ is well approximated by

$$\Lambda(k) = \Lambda(k_0) + \frac{1}{2}\Lambda''(k_0)(k-k_0)^2 \qquad \text{(M5)}$$

so that $\Pi(k)$ is approximately normally distributed, and, e.g., 95% confidence intervals correspond to values of $k$ for which $\Lambda(k_0) - \Lambda(k) \approx 1.92$.

### Selection between hypotheses

We add two procedures to systematically improve the hypotheses (see Fig. 1): one, in the spirit of [55] (compare also [56]), avoids overfitting by removing irrelevant parameters, another allows to include overseen aspects into the hypothesis.

Consider overfitting first. Adding parameters to the model is beneficial only if better description of the data, i.e., information content of the discrepancy between the model and the data, outweighs the increase in the complexity of the model, i.e., its information content, which can be approximated as $\mu l$, where $l$ is the number of parameters used and $\mu$ is information content per parameter. Then, having a parameter in the model is beneficial only if its presence results in the increase of the loss function

$$\Delta L > \mu \qquad \text{(M6)}$$

(we choose $\mu = \log 100$ in this work). This approach is equivalent to optimization over an ensemble of models, with parameter $\mu$ playing the role of chemical potential coupled with the number of parameters. In practice, we start with a hypothesis with maximal set of parameters and try excluding them one by one, each time checking if $\Delta L$ is larger than $\mu$. We repeat the procedure consequentially until no more parameters can be excluded.

Finally, after the model is pruned of irrelevant parameters, it produces a list of outliers, i.e. the cities with smallest p-values. We study these outliers qualitatively, search for possible explanations of their behavior, refine the hypothesis accordingly, and repeat the fitting and parameter removal procedure. This feedback loop is repeated until we are not able to identify any new relevant attributes, thus inserting a quantitative modelling step into the usually qualitative cycle of hermeneutic interpretation.

**Dataset and data preparation**

We use the corpus of the Soviet Newsreel ``Novosti Dnya'' (News of the Day) sourced from the Russian footage archive Net-Film[57] with owners' permission. It consists of over 1700 short (approximately 10-minute) films that is almost complete for 1954--1992 (excluding 1965) with some additional issues from 1944-53. In 1954-86 the issues are weekly, and in 1987-91 bi-weekly. Most newsreels contain multiple short news stories, although there are occasional single-topic issues dedicated to major political events (see [24, 46] for more details).

The corpus metadata includes story outlines in Russian, which we cleaned, split into stories (12,707 overall) and used for further analysis (see [46] for the details of data preparation). Approximately 97.5% of the stories are from the period between 1954 and 1986, the median being 1968.

Cities are included in the list of cities of interest if they exceed preset population levels [46] (for USSR cities we use 1959, 1970 and 1979 census data[58], for cities outside USSR we mostly use the UN Population Division data for 1970 [59]). The mentions of each city where manually classified by native Russian-speakers into 5 categories: (i) direct mention of a city and city-dwellers, (ii) mention of organizations and industrial enterprises located in the city and named after it, (iii) mentions of the region surrounding the city, and organizations located there, (iv) entities named after the city but located elsewhere, (v) homonyms and coincidences. In what follows, only mentions of type (i) and (ii) are considered.

**Acknowledgements:** This project was funded by the CUDAN ERA Chair project for Cultural Data Analytics, funded through the European Union's Horizon 2020 research and innovation program (Grant No. 810961). MT also acknowledges support from the Estonian Research Council (ETAG), Grant PRG 1059.

**Author contributions:**

Conceptualization: MT, MO, KM, MM, MS

Methodology: MT

Data preparation and cleaning: MT, MO, KM

Data analysis: MT

Visualization: MT, KM, MS

Writing – original draft: MT, MO

Writing – review and editing: MT, MO, KM, MM, MS

**Competing interests:** Authors declare that they have no competing interests.

**Data and materials availability:** All data used for the analysis, as well as full logs of the optimization procedure and results of each model for every city of interest are available at https://github.com/thummm/soviet_newsreels/. Correspondence and requests for materials to be addressed to MT, thumm.m@gmail.com.

**Code availability:** Optimization of all the models described in the text has been done using custom made code in Wolfram Mathematica 13.1. The code is preserved and will be made publicly available at https://github.com/thummm/soviet_newsreels/ upon publication.

## Figure legends

,

**Figure 1. The workflow pipeline of the suggested procedure** to extract information on media representation of cities. The black arrows correspond to the flow of data. The green arrow denotes classification of hypothetical parameters into relevant and irrelevant according to a predetermined information theoretic criterion. The red arrow signifies the feedback loop, i.e. the systematic refinement of the hypothesis based on the qualitative study of model outliers.

**Figure 2. Cities of interest on the map of the USSR.** Cities with population exceeding 0.03% of the USSR population and their mentions vs. expected from population-only model. Significantly ($p < 0.05$) over- and under-represented cities, insignificantly ($0.5 > p > 0.05$) over- and under-represented cities and cities which are mentioned roughly as expected ($p > 0.5$) are shown with cyan, red, grey-cyan, grey-pink and grey circles, respectively. Cities in Moscow and Donbas regions are shown in smaller circles to reduce their overlap.

**Figure 3. Observed city mentions vs expectation from population-only and full models.** City mentions vs. (A) population of the cities and (B) prediction of the full model, which accounts for population, geographical regions and city specialization for all cities with population above 0.03% of the population of the USSR. Cities mentioned zero times in the dataset are shown in black, out of scale. The red straight lines correspond to ideal correspondence with model and observation (power law regression (1) in panel (A), identity in panel (B)). Dashed and dotted lines correspond to deviations with p=0.05 (dashed) and p=0.001 (dotted). Note that number of big outliers is much smaller in the full model (cf. cities outlined with black circles).

**Figure 4. Overview of models explaining Soviet city representation.** Top left: seed regions used to initiate optimization. Borders of union-level republics and borders of subregions inside Kazakhstan, Russia and Ukraine are shown in blue and red, respectively. Top right: relevant regions according to the geolocation model, overmentioned regions shown in gradations of blue, underrepresented – in gradations of yellow. Bottom right: relevant city specializations. Bottom left: relevant regions according to the full model, overmentioned regions shown in gradations of green, underrepresented – in gradations of pink. See Table 2 for the values of regional and specializational boost factors.

# Tables

| City | Mentions | Pop, mln | City | Mentions | Pop, mln |
|---|---|---|---|---|---|
| Moscow | 2831 | 7.06 | Tokyo | 16 | 23.3 |
| St. Petersburg | 339 | 3.95 | New York | 29 | 16.2 |
| Kyiv | 95 | 1.63 | Osaka | 3 | 15.3 |
| Tashkent | 45 | 1.38 | Mexico | 6 | 8.83 |
| Baku | 38 | 1.26 | Buenos Aires | 3 | 8.42 |
| Kharkiv | 43 | 1.22 | Los Angeles | 0 | 8.38 |
| Nizhny Novgorod | 45 | 1.18 | Paris | 39 | 8.21 |
| … | … | … | … | … | … |
| Minsk | 72 | 0.92 | Berlin | 62 | 3.21 |
| Volgograd | 62 | 0.86 | Warsaw | 64 | 1.30 |
| Riga | 73 | 0.73 | Prague | 51 | 1.08 |

**Table 1. Number of mentions of selected cities** inside (left) and outside (right) the USSR. Population values are for 1970, the data for the top 7 cities by population, and 3 most mentioned ones outside the top 7 is shown.

| **Parameter** | Specialization model | Full model |
|---|---|---|
| Residual ***c*** (for the full model: the value outside 6 special regions mentioned below) | 1.42 (1.33…1.57) | 1.66 (1.50…1.88) |
| Scaling exponent ***a*** | 1.14 (1.09…1.17) | 1.18 (1.13…1.22) |
| **Boosts due to city specialization** $s_\alpha$ | | |
| Union-level capital | 1.65 (1.48…1.81) | 2.00 (1.77…2.22) |
| Autonomy capital (inside Russia proper) | 0.75 (0.61…0.88) | 0.74 (0.61…0.89) |
| Seaside (Black, Baltic or Pacific) | 2.38 (2.15…2.62) | 1.91 (1.68…2.10) |
| Hydroelectricity ( > 2 GW) | 2.46 (2.08…2.80) | 2.14 (1.77…2.50) |
| Steelworks | 1.53 (1.32…1.72) | 1.84 (1.54…2.14) |
| Non-ferrous metallurgy | 1.50 (1.27…1.79) | 1.55 (1.22…1.92) |
| Coal mining | 0.53 (0.43…0.65) | 0.62 (0.50…0.75) |
| **Regional boosts** $k_\alpha$ **(regional multipliers to *c*)** | | |
| Up to 250 km from Moscow | | 1.38 (1.18…1.59) |
| North Kazakhstan | | 1.38 (1.09…1.71) |
| Center, mid- and lower-mid Volga | | 0.78 (0.67…0.87) |
| East Urals, West Siberia | | 0.65 (0.55…0.77) |
| Central and Southern Ukraine, Donbas, Rostov | | 0.58 (0.50…0.64) |
| Central Asia, Armenia, Azerbaijan, Southern and Western Kazakhstan, West Urals | | 0.46 (0.39…0.53) |

**Table 2. Parameters of the specialization and full models.** Optimal values and 95% confidence intervals (in brackets) of the parameters of the specialization and full models, obtained by the minimization procedure. See bottom-right panel of Figure 4 for the exact shape of the relevant regions.

| **Parameter** | **Value** |
|---|---|
| Residual $\boldsymbol{c}$ (number of mentions of a city with 1 mln population) | 0.14 (0.12…0.20) |
| Scaling exponent $\boldsymbol{a}$ | 1.26 (1.14…1.32) |
| **Regional boosts $\boldsymbol{k_\alpha}$ (regional multipliers to $\boldsymbol{c}$)** | |
| Socialist I: Albania, Bulgaria, Czechoslovakia, Mongolia | 64 (43…82) |
| Socialist II: East Germany, Hungary, Poland | 37 (27…45) |
| Socialist III: Cuba, North Korea, Romania, Vietnam, Yugoslavia | 16 (11…21) |
| Capitalist I: Austria, Finland | 74 (50…97) |
| Capitalist II: the rest of Europe | 4.7 (3.5…5.9) |
| Capitalist III: Australia, Canada, USA | 2.5 (1.8…3.2) |

**Table 3. Parameters of the foreign cities model.** Optimal values and 95% confidence intervals (in brackets), obtained by the minimization procedure. Notably, all mentions of Tirana (Albania) are before the Soviet-Albanian split of late 1950s.

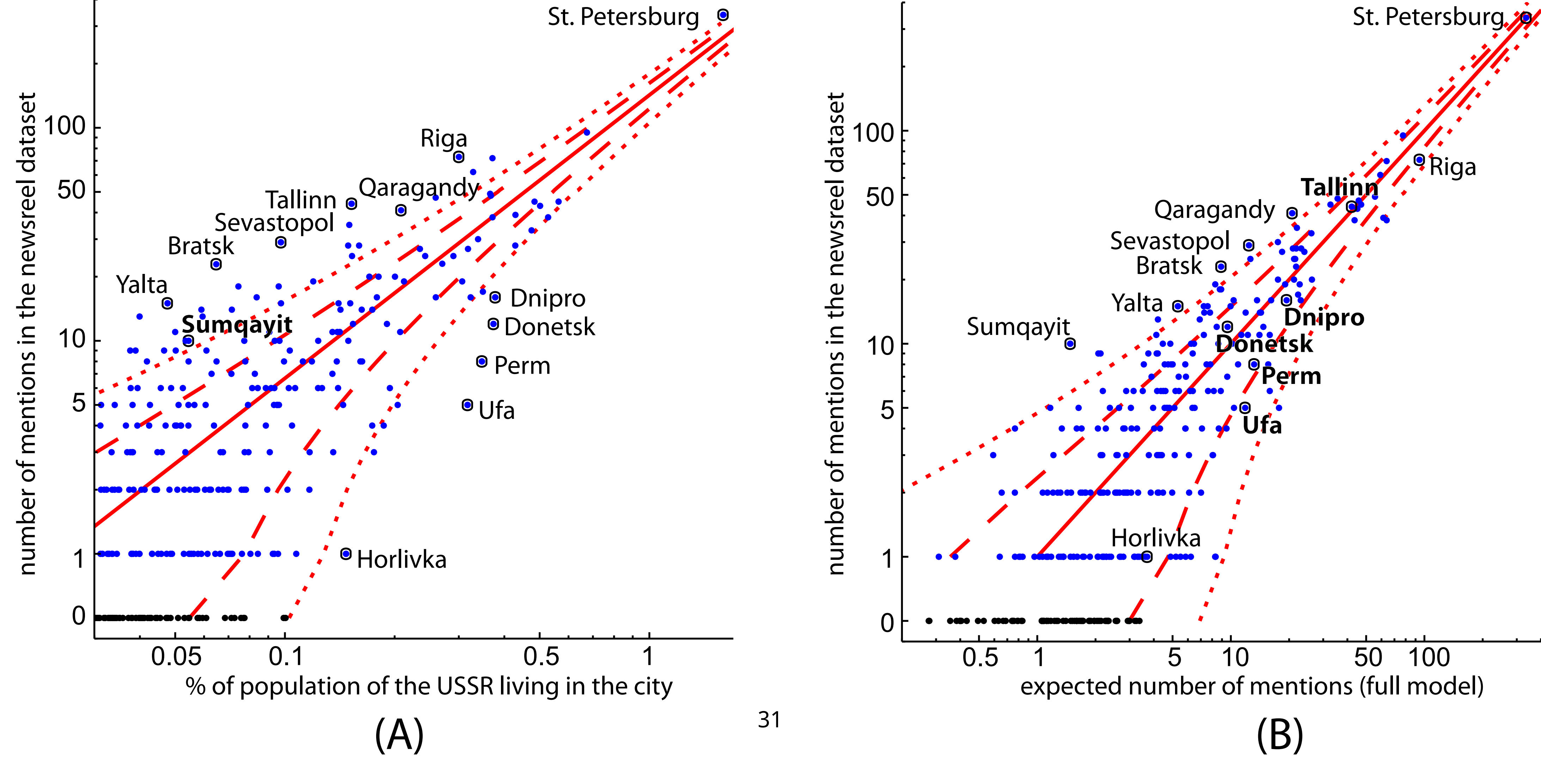

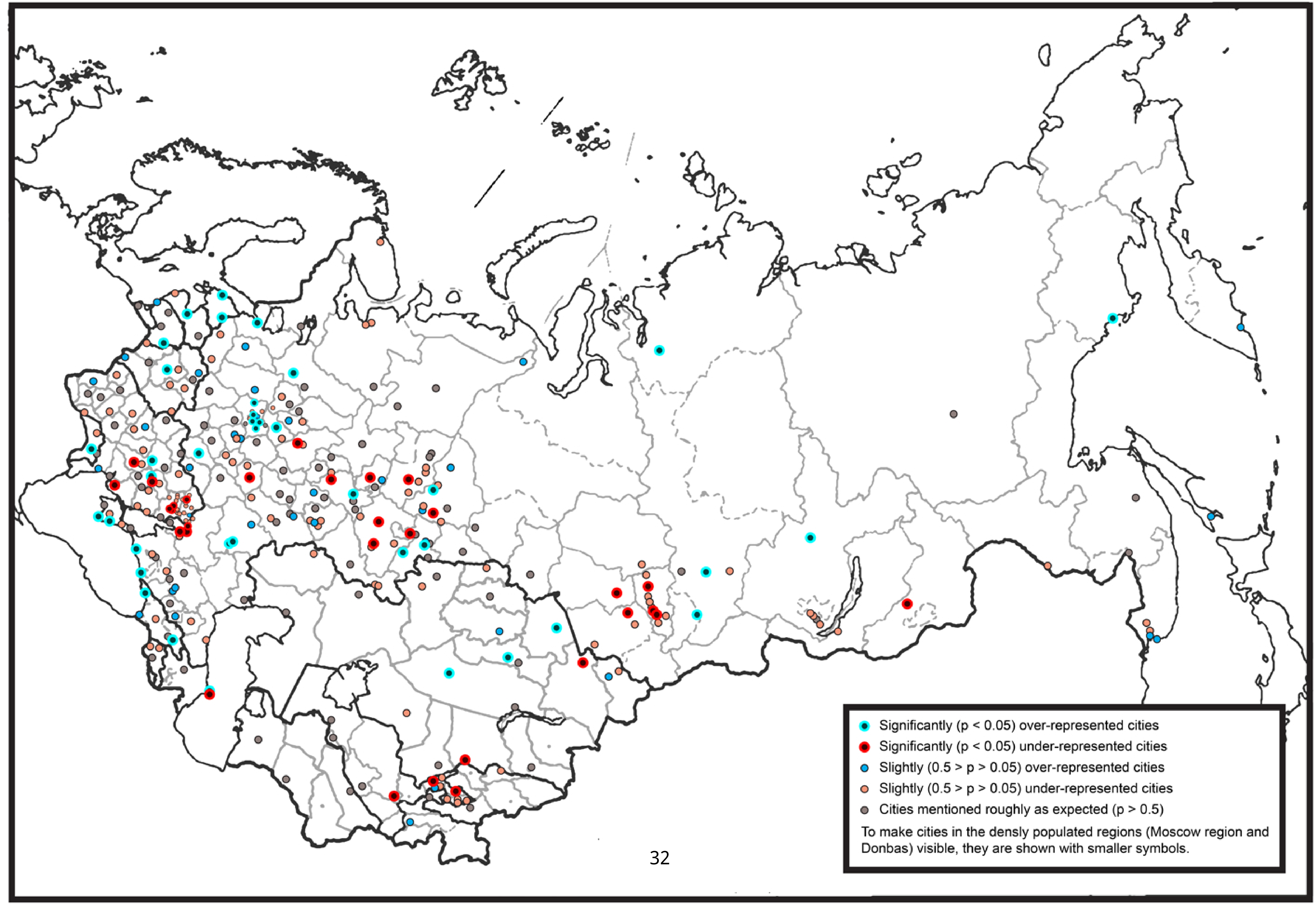

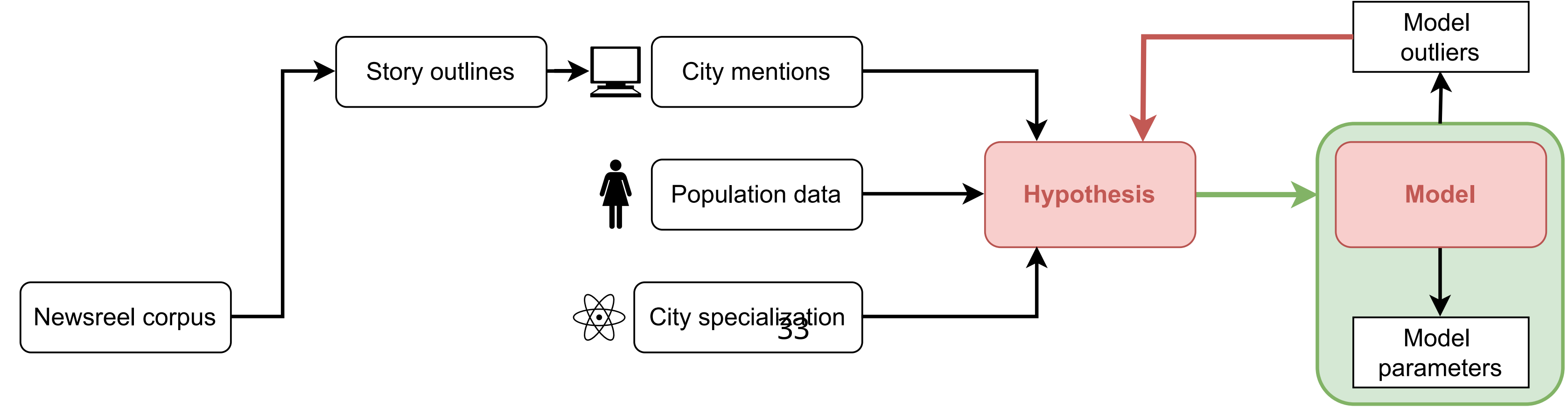

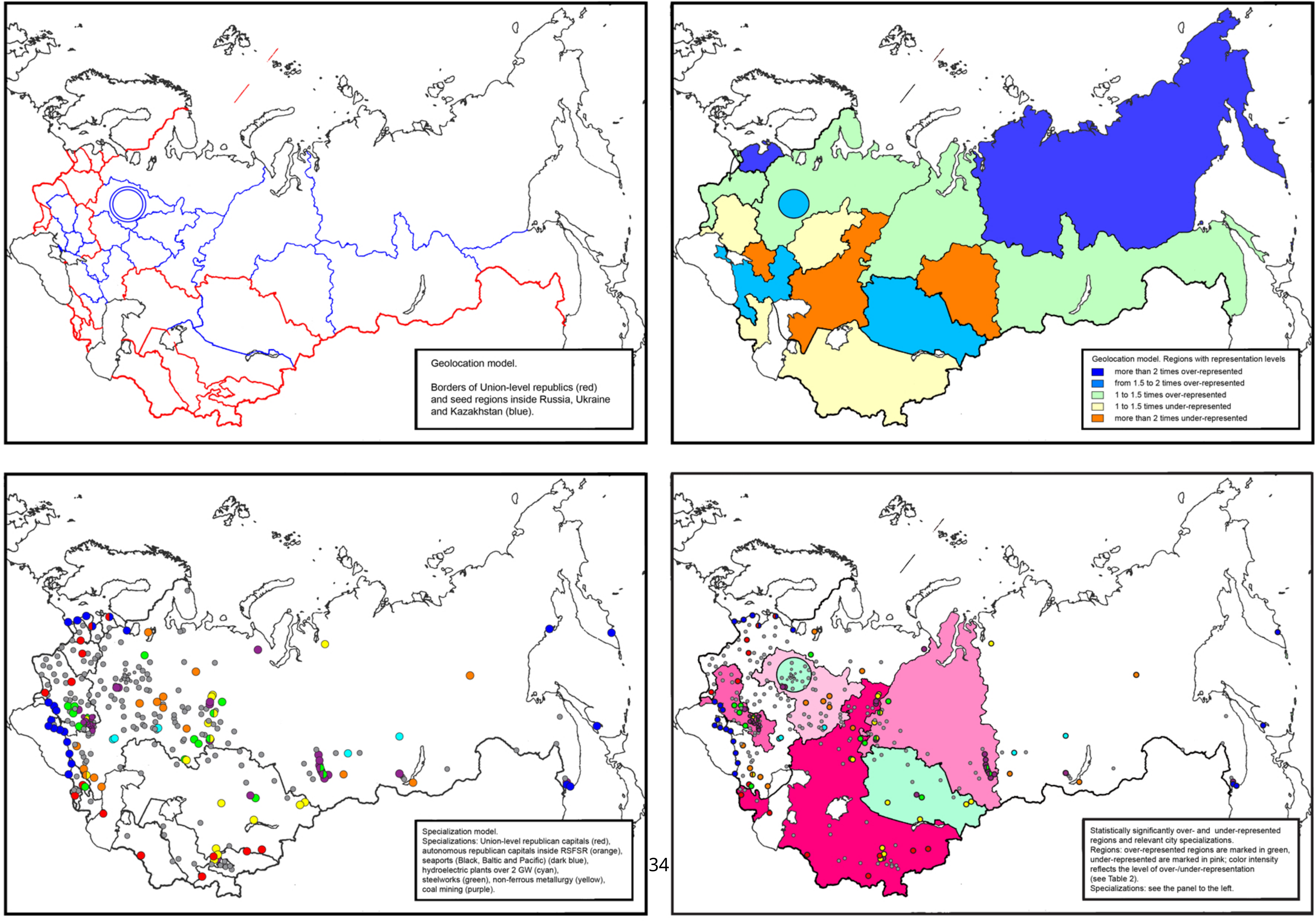
Geolocation model.
Borders of Union-level republics (red) and seed regions inside Russia, Ukraine and Kazakhstan (blue).
Geolocation model. Regions with representation levels
more than 2 times over-represented
from 1.5 to 2 times over-represented
1 to 1.5 times over-represented
1 to 1.5 times under-represented
more than 2 times under-represented
Specialization model.
Specializations: Union-level republican capitals (red), autonomous republican capitals inside RSFSR (orange), seaports (Black, Baltic and Pacific) (dark blue), hydroelectric plants over 2 GW (cyan), steelworks (green), non-ferrous metallurgy (yellow), coal mining (purple).
Statistically significantly over- and under-represented regions and relevant city specializations.
Regions: over-represented regions are marked in green, under-represented are marked in pink; color intensity reflects the level of over-/under-representation (see Table 2).
Specializations: see the panel to the left.

Supplementary Materials for

# City representation in the Soviet propaganda: quantifying biases of the Soviet worldview

Mikhail V. Tamm, Mila Oiva, Ksenia D. Mukhina, Mark Mets, Maximilian Schich

## Data preparation

### Dataset characterization

The corpus of Soviet Newsreel “News of the Day” (Новости дня / Хроника наших дней] was downloaded from Russian footage archive Net-Film[1] with permission of the owners, it was previously introduced and discussed in Ref [2]. The “News of the Day” journal was the main newsreel journal produced by the Central Studios of Documentary Film in Moscow. The corpus includes almost all issues of this newsreel from 1954 to January 1992 (except for the year 1965), as well as a few surviving issues from 1944 to 1953. Figure S1 illustrates the contents of two exemplary newsreels.

In Figure S2a the number of issues per year is presented. Starting from 1954 the newsreels have been saved systematically, and the newsreel production have peaked with 72 reels in 1954 and 65 in 1955. For thirty years, in 1956-1986 the usual annual number of newsreels was stable at 48-52 issues, meaning approximately one issue per week. Starting from 1987 the annual number of newsreels dropped to 26 issues (1 issue in 2 weeks). Overall, the corpus includes more than 1700 short films of usually 9-10 minutes length.

The films are complemented with metadata, including the information on the issue number, the crew, and the short outlines. The newsreels and metadata are in Russian; three members of the research team (MT, MO and KM) are fluent in Russian and thus were able to perform data cleaning, preparation and preliminary analysis.

Typically, each newsreel is split into several (usually 5-10) short news stories. These stories are typically well separated (e.g., by a black screen between them) and are topically unrelated. There is a small fraction (around 3%) of single-topic issues (year-end, celebration-related, dedicated to party congresses, etc.) which either consist of a single story or a sequence of very short stories (up to 15 in 10 minutes) filmed in different places but united by a single topic (e.g., "working women in the USSR"). Finally, the dataset includes 30 double issues, i.e., two consequential issues united into a single film on a single topic. These are dedicated mostly to big political events, 18 out of 30 double issues are in years 1990-91.

Stories and outlines

We use an outline of a story as an elementary unit of analysis. We mostly use the outlines available in the complementary metadata. We made an extensive random check and found that the outlines are of satisfactory quality, with a very small number of mistakes: the fraction of outlines with typos in place names was significantly below 5%, and we only once (out of several hundreds checked) been able to find a film outline in which one of the stories was missing. In the vast majority of cases the format of outlines allowed automatic splitting into stories. Exceptions where (i) around 1% of newsreels where there were typos in the numbering of stories within a newsreel which we corrected manually, (ii) around 3% of newsreels (most of them from years 1989-92) which had a different format of outlines: instead of a contents summary it included description of camera movement, wide shots vs close-ups, etc.; for these roughly 50 newsreels we have rewritten the outlines to match the format of the rest.

Overall the dataset consists of 12 707 story outlines (on average 7.5 per newsreel), in \fig{newsreel}b their distribution by year is presented, the full list of the outlines is provided in [3]. It is seen that the huge majority of the dataset (97.5%) corresponds to 1954-1986, i.e., the period between the death of Stalin and the early years of perestroika. Interestingly, the number of stories per newsreel issue trends down with time, especially after 1975. Median date of a story is 1968 and 50% of stories belong to the period 1960-76 with 25% dated before and 25% after this period.

The choice of outlines as a data source as opposed to using, the automatic transcripts of the narrator's voice is due to their much higher quality: at the time when data preparation was done the automatic transcript software for Russian language produced large number of mistakes, especially in the names of persons and geographic locations, which is essential for this work. This approach clearly has its limitations. For example, it excludes cases where the place of filming is not explicitly mentioned, and it takes no account for the screen time dedicated to a geographic location or to the related aspects of visual aesthetics[10]. Without doubt, the progress in AI technologies will soon make it possible to go beyond these limitations.

City population

For estimates of the city population we use the USSR censuses for Soviet cities and UN and (if needed) national data for foreign cities.

In the case of USSR population is an extremely important variable (see Fig. 3A of the main text), and USSR census is a relatively consistent and reliable dataset. As a proxy of the population we use an average fraction of population of the USSR living in a given city averaged over three censuses of 1959, 1970 and 1979 [4]. The list of cities of interest include all 309 cities with population more than 0.03% of the population of the USSR, except Moscow. For the purposes of models that include additional variables apart from the population one, we further enrich the list to make sure that 5 largest cities of each union-level republic is included. This increases the size of the dataset to 328 cities. It is done to avoid too small grouping of cities and contrast capitals of Union-level republics with non-capital cities of the same republics. We use population of cities ``including other urban dwellings answering to the city council'' since it correlates with the number of mentions slightly better than the population of city proper. Note, however, that large discrepancy between population of city proper, and population including other urban dwellings is especially common for coal-mining towns. As a result, their observed underrepresentation (see Table II of the main text and auxiliary tables) might be partly due to this decision.

Unfortunately, due to varying standards of the national statistical bodies, there is no equivalent universal dataset for population of the cities worldwide (note, however, that huge discrepancies between population of metropolitan areas and cities proper is less common in 1950s-70s than in the modern period). In the absence of such a dataset we use, wherever possible, the 1970 estimate from the 2018 World Urbanization Prospects Report of the UN Population Division[5]. For the cities, for which such estimate is not available, we use data from national statistical bodies. In case there is no data for 1970, we approximate population linearly between two closest censuses before and after 1970.

These complications do make the population figures for foreign cities somewhat ambiguous. However, we found that for foreign cities population plays much smaller role in determining city mentions than in the case of Soviet cities. Indeed, if for Soviet cities, according to the geography model, a city from the most popular region (North-East) is mentioned similarly as a 4.8 times larger city in the least popular region (West Urals), for foreign cities a city in the most popular region (Austria and Finland) is equivalent to a city 30 times larger from the least popular one (third world). We therefore expect that minor ambiguities in the population variable for cities from different countries are not particularly relevant.

The list of cities of interest initially consists of 135 cities with population above 1 mln in 1970 and is further enriched to allow for the fact that capitals, cities in Europe and in socialist countries are mentioned more frequently. To do that, we include all capitals and all cities in Europe and in China with population above 0.5 mln, European capitals and cities in non-European socialist countries with population above 0.25 mln, and all cities in European socialist countries, Austria and Finland with population above 0.1 mln. The resulting enriched dataset includes 310 cities. Of these only 113 are mentioned at least once, but recall that our approach allows to extract information from cities with zero mentions.

In order to roughly estimate the mentions of cities outside the aforementioned close lists we use slovnet[6], a Python library dedicated to analyzing Russian language, to extract named entities from the story outlines. By analyzing the output of slovnet we found that there are some places outside the cities of interest lists, which are mentioned extensively, including, for example, Tynda (the end point of Baikal-Amur railroad, an important construction project of 1970-80s), Mikhailovskoye (birthplace of Alexander Pushkin) and Zvezdny Gorodok (a place where Soviet astronauts were trained) inside the USSR, and Geneva (location of many important international negotiations) outside it.

Creating a full clean list of places mentioned in the dataset implies very significant manual work and is not needed from the point of view of the methodology presented here.

The task is especially daunting in the case of places inside the USSR, in part because they are mentioned more, in part because of the large number of places with coinciding names, places named after prominent communist politicians, which are easy to confuse with mentions of those politicians themselves and other entities (streets, plants, collective farms) named after them, etc. That is why we only produced this analysis for foreign cities. The results are summarized in table S1. Thus, the cities of interest constitute more than 60% of all foreign places mentioned in the dataset and contribute more than 90% to all mentions of foreign places.

City mentions

For each city in the list, we obtained and cleaned the corresponding list of mentions. In order not to miss any relevant mentions, for each city the story outlines were searched for matching substring(s) covering all possible Russian word forms derivative from the city name (these substrings were selected from Wictionary [7] and pymorphy library[8] and supplemented by the authors' knowledge of Russian grammar). For cities whose names names has changed during the Soviet period (Mariupol/Zhdanov, Volgograd/Stalingrad, Leningrad/Petrograd, etc) all forms of the name were checked.

The resulting lists of matches were classified manually into relevant and irrelevant mentions. This stage is reasonably fast for the dataset of this size (roughly 2 weeks of work) but is not scalable for larger datasets like, e.g. full corpora of TV news or newspapers for a period of similar length. However, (i) this work must be way easier for analytical languages like English, Chinese or French, (ii) there is strong evidence (see, e.g. [9]) that such tasks can now be automated with reasonably high precision using large language models.

One particular complication typical for the Soviet period is that in many cases multiple entities are named after the same prominent person, so that additional research is needed to disentangle them. One illustrative example is the difference between Gorky train line ("Горьковская железная дорога") and Gorky metro line ("Горьковская линия метро") in Moscow: both are ultimately named after the writer Maxim Gorky; however, the former is named after the city of Gorky (now Nizhny Novgorod) which is in turn named after the writer, while the latter is called after Gorky street in Moscow (which is named after the writer) and is unrelated to the city of Gorky.

We used the following classification of city mentions:

Type 1 - direct mention of the city as a location of filming or of city-dwellers;

Type 2 - mentions of entities (plants, universities, football teams, etc) located in the city and having city name or city-derivative adjective in their name (Moscow State University - Московский государственный университет, Dynamo Kyiv - Киевское Динамо, ...);

Type 3 - mentions of the area surrounding the city, which can take the form of mention of the city name with specification "рядом с" (near), "неподалеку от" (not far from), etc., administrative divisions (oblasts, etc) named after their center city, as well as informal geolocation names like "Подмосковье" (Moscow region), "Рижское взморье" (Riga seacost), etc.;

Type 4 - mention of the objects and entities named after the city but not located in or near it, like Warsaw pact or Paris commune shoe factory;

Type 5 - irrelevant: there is an automatic match but it is a coincidence, due to random homonymy or similar origin of the name like in the Gorky example above.

Occasionally, an outline of a story mentions a single city multiple times. Such a multi-mention is counted as a single mention and is assigned the type with the smallest number. For example, a phrase “В Варшавском аэропорту прошла торжественная встреча делегаций, прибывших в Варшаву на саммит стран Варшавского договора” (A ceremonial reception for the delegations arriving in Warsaw for the Warsaw Pact countries' summit took place at Warsaw Airport), which includes Type 1 mention (прибывших в Варшаву) includes type 1 mention (Warsaw per se), type 2 mention (Warsaw airport) and type 4 mention (Warsaw pact), and is counted as a single type 1 mention.

All mentions of the cities in the cities of interest list are manually classified into these 5 types. For consistency, all annotations used in the further analysis, are done by MT. To check the reliability of human annotation two other Russian-speaking members of the team (MO and KM) made test annotation of 407 story outlines related to 12 selected cities (Baku, Izhevsk, Helsinki, Kaunas, Kursk, Lviv, Novgorod, Paris, Ryazan, Sofia, Tomsk, Tula) according to the following instruction:

***

Annotation instruction

For each story in the list separately

i) Find all mentions of the city and city-named entities in the text of the outline.

ii) if the city or city dwellers are mentioned directly, classify as 1 ["Москвичи вышли на парад", "Новосибирск. Ловля лосося", "на шоссе Киев-Краснодар...", "матч Динамо (Тбилиси)"]

iii) if not already classified, but there is an entity mentioned which is named after the city and located in it, classify as 2 ["Горьковский автозавод", "Московский кинофестиваль", "Бакинский ансамбль народных танцев"]

iv) if not already classified but there is a mention of the region centered in the city, or the vicinity of the city, or of the entity named after the region, classify as 3 ["уборка свеклы в колхозах Винницкой области", "соревнования под Красноярском", "Калининская атомная электростанция в Удомле"]

v) if not already classified but there is a mention of the entity named after the city but located elsewhere/nowhere, classify as 4 ["Казанский вокзал в Москве", "Фабрика имени Парижской коммуны", "страны Варшавского договора"].

vi) else, if mention is simply homonymy or mistake, classify as 5.

If possible, try to figure out where the mentioned entities were located. If in doubt or borderline classify explicitly agricultural entities ("Кишиневский экспериментальный совхоз") as located in the vicinity of the city (i.e., classify as 3), and all the other (industrial, cultural, etc) ones ("Сталинградская ГЭС") as located within a city (i.e., classify as 2).

When classifying, take into account, not only where the event is taking place but also where the mentioned entity is located: "матч Динамо (Киев) в Тбилиси", "выступление шахтера шахты X (Ленинск-Кузнецкий) на всесоюзной партийной конференции в Москве", "На Ленинградский завод моторов закончено производство 218й турбины для Красноярской ГЭС" are counted as mentions of Kyiv, Leninsk-Kuznetsky and Krasnoyarsk, respectively (they are also counted, of course, as mentions of Tbilisi, Moscow and St Petersburg).

***

The full results of this annotation are available at [3]. Table S1 summarizes the most important results, showing that both precision and recall of the annotation used (if the result of the alternative annotators is considered a ground truth) is around 95%. A more detailed analysis of discrepancies shows that they are mostly due to human error (more or less equally distributed between annotators) and partly to different treatment of borderline cases. The tables of mentions for these representative cities, marked-up by two annotators independently, are available in the supplementary Annotation.Comparison.zip Archive

**Detailed results of the models**

Together with this supplementary text we provide two supplementary tables in the .xslx format, containing the detailed information on the run of all studied models for the Soviet and foreign cities [3]. Below we give the detailed outline of the structure of these files and the information contained in them. We also provide multiple comments on various aspects of the results.

<u>Soviet cities models</u>

I. *Raw data on mentions and population*. Master table contains full information on the contemporary Cyrillic name(s) of the cities in the cities of interest list, their population at each of the three censuses, and the number of mentions of each city in the dataset.

II. *Results for the population-only model*. Pop_only_pval table contains the results of the population-only model, including comparison of actual mentions of each city with corresponding predicted mentions, and individual p-value of each city. Thus, 24 cities are over-mentioned with $p < 0.001$ and 6 cities are similarly undermentioned. Tallinn, Bratsk, Riga, Sevastopol, Yalta, Rustavi, Vilnius, Cherepovets, Minsk and Volzhsky form the top 10 of most significantly overmnentioned cities. Conversely, Ufa, Perm, Donetsk, Dnipro, Horlivka, Kemerovo, Kazan, Novokuznetsk, Barnaul and Baku are the top 10 most significantly undermentioned ones.

*The role of censoring.* We checked how different choices in the level of censoring the cities by population influence the results of the model. The corresponding results are provided in Table S3. Clearly, although including more cities reduces the confidence intervals for the parameters, the confidence intervals strongly overlap for censuring at 0.03%, 0.05% and 0.1% of the population of the USSR.

*The influence of Moscow.* As mentioned in the main text, two properties of Moscow – being the capital of the USSR and being the host city of the "Novosti Dnya" newsreel production – make it incomparable to other cities of the USSR. As a result, Moscow is mentioned roughly 5 times more than expected from population only model. Therefore, it is not surprising that it inclusion shifts the scaling data dramatically (see Table S3): the loss function puts a lot of weight on fitting this one big outlier to the detriment of the fitting the rest of the data. On the other hand, the only imperfect comparison available is to the capitals of the foreign cities, where we found that capital effects can be estimated by replacing the city population by the geometric mean of the populations of the city and the corresponding country. If this renormalized population is used for Moscow, it turns out that it is in fact undermentioned by a factor of roughly 2 as compared to the prediction of the population-only model, which might indicate that capital effect work differently here and/or that significant fraction of stories are located in Moscow by default without explicit mention in the outlines. In any case, Moscow is a completely unique case and we exclude it from further consideration.

*The influence of St. Petersburg on the fit.* After Moscow is excluded, St. Petersburg is the second significant outlier both in terms of population and in terms of mentions: it is 2.2 times larger than second largest city in the dataset (Kyiv). Since it does not have the unique properties of Moscow, we keep it in the dataset, but check how much this single point influences the results of the fitting. We found that, indeed, there are some minor but notable changes in the results of the model optimization over the whole dataset and over the same dataset but without St. Petersburg (see the three last columns of the Pop_only_pval table). First, the optimal value of the scaling exponent is slightly smaller $a = 1.24 \pm 0.05$ instead of $a = 1.33 \pm 0.04$ for the full dataset, which is borderline significant (see Table S3). Second, the ordering of the most over- and under-mentioned cities slightly changes. In particular, St. Petersburg and Volgograd replace Cherepovets and Volzhsky in the list of most overmentioned cities, with St. Petersburg becoming the most significantly overmentioned one. In turn, Dzerzhinsk and Chita replace Kazan and Baku in the list of the most undermentioned ones. These changes are, however, relatively minor (except when discussing St. Petersburg itself). Therefore, we decided to keep the whole sample. Note nevertheless that results for St. Petersburg should be interpreted with a certain caution. Moreover, we have checked that if omission of any other city from the dataset does not change the results in a statistically significant way.

*Time evolution of the population-only model.* The data we study spans several decades of Soviet history. It is natural to ask how much the observed patterns of mentioning cities change throughout this period. Our ability to study this is somewhat limited due to the sparseness of the data. However, we provide here the results of the population-only model run on the data from three eleven-year periods: 1954-64, 1966-76 and 1977-87 (recall that 1965 is missing from the dataset, and more than 97% correspond to the 1954-86 interval). We use the population data from the 1959, 1970 and 1979 censuses, respectively, as a measure of city population, and use 0.05% population cut-off for the first two periods and 0.06% for the third, so that there are no cities above the cut-off which are not included in our 328-city dataset. Note that using the whole dataset without cut-off would have been methodologically wrong. For example, cities, which are small in the earliest period but subsequently become large enough to be included in the dataset do not form a representative sample of small cities.

The scatters plots of mentions versus population for each period are presented in Figure S3 and the parameters of corresponding models are summarized in Table S3. There are several important observations to be made. First, the overall number of mentions systematically decreases with time in agreement with the decreasing number of stories per year (compare Figure S2B). Second, for each period separately the number of mentions does scale with population size as predicted by the population model. In all cases the scaling exponents are above one with high confidence, indicating the presence of agglomeration effects. However, the scaling exponent trends down with time, i.e. in later period the distribution of mentions becomes less skewed towards larger cities. Third, on a single-city level there exist multiple different scenarios. Some of the most "popular" cities, e.g., Sevastopol and Tallinn, are overmentioned throughout each period separately. Mentions of some others, e.g., Krasnoyarsk, Qaragandy, Vladimir, are more localized in time (in case of Krasnoyarsk this is clearly connected to the construction of Krasnoyarsk hydroelectric dam). Fourth, the most dramatic change between the first period and the later two is related to the status of Kyiv. Indeed, in 1954-64 Kyiv is clearly the third most important city in the USSR hierarchy: it is mentioned significantly more than population-based expectation and has almost double the number of mentions of the fourth-most-mentioned city (which, interestingly, is Odesa, i.e., another Ukrainian city). Conversely, both in the 1966-76 and in the 1977-87 periods Kyiv is mentioned less than expected based on its population, and, despite remaining the third largest city in the USSR, is mentioned less than some smaller cities. The mentions of Odesa drop even more dramatically. One possible explanation for this change might be related to importance of Ukraine and Kyiv. Interestingly, this change coincides to a well-known shift from promotion of Ukraine as second-most-important republic of the USSR during N. Khruschev era to comparative neglect and insidious Russification in the later period [11,12].

III. *Results of geography, specialization and full models.* For each of these three models we provide four tables, specifying

(i) the list of variables used, including population, flags designating that a city belongs to a certain geographic group, and flags designating specializations present in the cities factors (sheets Geo_variables, Spec_variables and Full_variables).

(ii) log of the optimization process: which merges of geographical regions (omissions of the specialization variables) where attempted in which particular order, what were the results of loss function optimization, and whether attempts where accepted or not (sheets Geo_clustering_log, Spec_clustering_log and Full_clustering_log);

(iii) table of the resulting values of parameters and their confidence interval in the final version of the model, including the lists of optimized geographical regions, their composition, and corresponding boost factors (sheets Geo_confidence, Spec_confidence and Full_confidence);

(iv) values of actual and predicted numbers of mentions for each city, and corresponding p-values (sheets Geo_expectations, Spec_expectations and Full_expectations).

Apart from that, we provide two summary tables, specifying

(i) the list of seed geographical regions, their definitions, and which macro-regions they are allocated to by the optimized geography model and by the optimized full model (sheet Seed_regions);

(ii) the list of specializations studied, and whether they are statistically significant (sheet Specializations).

*Seed specializations and choice between them.* The initial list of specializations is provided in table S5 for a quick reference. Generally, we start with feeding into the model a wide set of variables, compatible with several alternative hypotheses, and then let the optimization evolve and choose one option out of many. Below we discuss three particular instances of this approach.

*Sub-republican autonomies and administrative units*. We start with distinguishing 4 classes of cities: capitals of autonomous republics inside and outside Russia proper, capitals of non-national sub-republican units (oblasts and krai's) and cities located inside autonomous republics but having no capital status. The model optimization process algorithm attempts to both (i) merge these classes of cities in different combination and (ii) discard them (i.e., essentially merge the classes with a "dummy class" of cities with no administrative function and located outside autonomies). In this case the optimization resulted in discarding all classes except for capitals of autonomous republics inside Russia proper, which turned out to statistically significantly reduce the representation. Note, however, that the size of the "capitals of autonomous republics outside Russia proper" class is very small (just 3 cities: Batumi, Sukhumi, and Nukus), making the inference in this case somewhat less reliable.

*Ports and recreation cities.* We start with 6 classes for port cities located ashore of various masses of water (Arctic and Pacific oceans, Azov, Baltic, Black and Caspian seas). We also introduce a class of cities specializing in recreation in order to check the hypothesis that predominantly recreational cities (e.g., Sochi, Yalta, Jurmala) are represented differently than predominantly military or trade ports (e.g., Sevastopol, Novorossiysk, Kaliningrad). Once again, in the model optimization state the city classes corresponding to the shores of different seas might be either merged or discarded, and the "recreation" class can be either preserved (meaning that there is statistically significant difference between recreational and non-recreational cities) or discarded. The result of optimization in this case is a bit unexpected: it turns out that there are two significantly different classes of seas: overrepresented Black, Baltic and Pacific on one side, and not overrepresented Arctic, Azov and Caspian on the other. This difference might possibly be rationalized by noting that the first set of seas is relatively more “outward looking” (that is, related to international transportation, international relations and corresponding history) than the second. Moreover, there is no significant difference between recreational seaside cities and military/trade ports.

*Hydroelectricity.* We started with two classes of cities with hydroelectric power dams, one corresponding to huge dams with power above 2 GW and medium-sized dams of 0.5-2 GW. It turned out that only huge dams lead to a statistically significant increase in representation. Notably, all 4 dams in question were built during the studied period, and it is mostly the construction stage that is being covered in the newsreels. However, the same is true for most of the medium-sized dams and leads, in their case, to no observed representation effect.

*Validation of specializations.* The industrial specializations, which we found relevant, particularly hydroelectricity and metallurgy, is well-known and reflected in Soviet culture on multiple levels, from heroic Komsomol songs to E. Evtushenko’s flagship epic poem “Bratskaya ges” (“The Bratsk Station") to sarcastic mentions in the openly anti-Soviet sources, like in this famous song by Y. Aleshkovsky:

И пусть в тайге придётся сдохнуть мне,
Я верю: будет чугуна и стали
На душу населения вполне.

(“And even if I have to kick off in taiga, I believe: there will be enough cast iron and steel per capita”).

However, it is interesting to check that this is not a coincidence and the specializations which the model finds to be relevant are indeed represented in the data. We made a direct check of the topics of stories mentioning 7 representative cities (Odesa, Krasnoyarsk, Tbilisi, Cherepovets, Kazan, Oskemen and Donetsk) and counted the stories directly related to the relevant city specializations, the results are provided in Table S6. In most cases (e.g., mentions of Cherepovets steelworks) the counting is very straightforward, except for the number of “capital status” related stories are an estimate from below, as we counted only the most unquestionable ones (stories directly mentioning Georgia in case of Tbilisi and Tatarstan instead of Kazan). The table shows that indeed the specialization-related stories contribute quite significantly to mentions of corresponding cities. Moreover, the fraction of such stories is seemingly higher for the representation-boosting specializations.

*Representation of regions in the full model.* We find (see Figure 4D and Table 2 of the main text) six contingent regions, in which representation significantly differs from that in the rest of the country. Importantly, there are more deviations down then up from this default (“rest of the country”) level, i.e. this level itself is slightly (about 20%) elevated above the average over the whole of USSR. We relate overexpression of Moscow region to its geographical accessibility and Northern Kazakhstan to its importance in the virginlands reclaiming narrative. The slightly elevated “rest of the country“ region can be split into several groups of locations with different rationale for importance. We relate interest in Eastern Siberia and Far East with the exoticity of those places and narrative of expansion into wild lands („stroyki kommunizma“), in the South (Northern Caucasus, Lower Volga, Georgia and Blacksoil region) with its better climate and recreational attractiveness. Western part of the USSR (Baltic coast, Belarus, Moldova and Western Ukraine) seems to be of importance because of general Eurocentric bias of the Soviet woldview. This bias simultaneously explains the systematic neglect of the South-Eastern republics of the USSR (Central Asia, Armenia, Azerbaijan and Kazakhstan, except for its Russian-speaking North). Central parts of Russia proper (West Urals, and, to a lesser extent, Middle Volga, East Urals and West Siberia) seem to suffer from what we call “neglect of intermediate situations”: these parts are quite far away from Moscow, but still not virgin and exotic enough to warrant additional interest.

*Ukraine.* The most puzzling and interesting phenomenon is a very significant underrepresentation of the region covering most of Central and Eastern Ukraine, as well as Rostov region of Russia. Without doubt, the study of the role of this region in the Soviet worldview and its development in time (note the drastic fall in mentions of Kyiv and Odesa with time, see Fig. S3) is of extreme interest and importance, especially in the view of recent Russian aggression against Ukraine. Here we formulate a hypothesis about possible explanation. We conjecture that underrepresentation of Eastern Ukraine and Russia-Ukraine borderlands might be another manifestation of the "neglect of intermediate situations" pattern. The population of these regions was mixed, and identities of its residents formed a continuum spectrum from purely Russian to purely Ukrainian, including people speaking Russian but self-identifying as Ukrainian, bilinguals, speakers of Russian-Ukrainian pidgin language ("Surzhyk"), etc. This complexity resulted in Eastern Ukraine (except, maybe, purely Ukrainian-speaking Western part) to fall in-between of the standard Soviet nomenclature of nationalities. In turn, Central and Southern Ukraine (i.e., most notably Kyiv and Odesa) should have been seen as more properly-Ukrainian in the 1950s and 1960s, where it was [11,12] ideologically fashionable to celebrate Ukraine-ness as something distinct (although inseparably united with Russia), and as more in-between (i.e., similar to Eastern Ukraine) in 1970s and 1980s, the age of tacit Russification of Ukraine.

That is to say, we suggest that for Soviet ideologues might have felt that Eastern (and later also Central and Southern) Ukraine are perplexingly "neither fully Eastern European nor fully Russian" and, as such, better left without discussion.

IV. *Model comparison*. Finally, we provide a table with comparative summary of the models, which includes information on the number of outliers with *p*-values below 0.0001, 0.001, 0.01 and 0.05, as well as $R^2$ and normalized deviation $\langle\sigma\rangle$ defined as

$$\langle\sigma\rangle = \left(\frac{1}{K}\sum_{i}\frac{(n_i - m_i)^2}{m_i}\right)^{1/2} \tag{S1}$$

where $n_i$ is the number of mentions of $i$-th city, $m_i$ is the corresponding expected number, and $K$ is the total number of cities in the dataset. Note that for a set of Poisson random variables with expected values $\{m_i\}$ the value of $\langle\sigma\rangle$ is expected to converge to 1. Thus, $\langle\sigma\rangle$ has the meaning of "how large are the observed deviations from expectations as compared to the situation when such deviations are due purely to random noise".

It can be seen that on all metrics both geography and specialization models are a significant improvement on the population-only model, while full model is a significant improvement on them both. On balance, it can be argued that geography model explains the data slightly better than specialization one, however note that geography model has 16 relevant parameters (scaling exponent and expression levels in 15 regions), while specialization model has only 9 (scaling exponent, residual expression level, and boost factors for 7 relevant specializations). Meanwhile, it is striking that full model has a significantly larger explanatory power than the geography one despite having just 15 relevant parameters.

In terms of particular metrics, note that switch from population-only to full model allows to eliminate large outliers almost completely (from 19 to 3 cities with $p < 0.0001$) and to reduce the number of moderate outliers from 69 cities with $p < 0.05$ for the population-only model to 41 for the full model (note that in the dataset of $K$=328 cities one expects roughly 16 such outliers for purely random reasons, so the number of excess outliers is reduced by a factor of 2). Other natural metrics, such as $(1 - R^2)$ and $(\langle\sigma\rangle - 1)$ tell the same story: the full model allows to explain 50%-60% of variation unexplained by the population-only model.

Foreign cities model

The table with the results of the foreign cities model has a similar structure. It contains (i) the master list of the cities of interest with their population, and associated variables (flag indicating the city is a capital, population of the country, geographical location), all populations used are as of 1970, with UN Population Division 2018 World Urbanization Report being the main source of data, and national census authority data used in the cases a city is absent from it;

(ii) the list of seed geographical areas used, and their assumed proximity (i.e., for which areas merger was assumed possible); note that (i) contrary to the Soviet cities model proximity here is understood politically rather than geographically, i.e., socialist countries form a complete graph in terms of proximity, Australia and Canada are connected, etc.;

(iii) model optimization log (i.e., sequence of simplifications attempted and whether they were accepted or not);

(iv) model optimization result, with values of all parameters and corresponding confidence intervals;

(v) model expectation for individual cities vs actual numbers of mentions, and corresponding p-values.

*Seed geographical areas.* The choice of initial geographical areas, as well as area-dependent censoring of city population is data-driven. The seed areas include, separately, all 13 countries recognized as “socialist” in contemporary Soviet discourse (both Comcon and non-Comcon); Finland and Austria, whose high representation has been observed in the data; and USA, Canada, Japan and Australia, for which we hypothesized that their representation might be different from neighbouring countries. The rest of the world was split on continental level into Africa, Asia, Europe and Latin America.

*Capital status.* The way the formula

$$\log m_{\mathrm{F},i} = \log c + a\left(\log P_i + \frac{1}{2} I_{i,cap} \log\frac{P_{i,c}}{P_i}\right) + \sum_\alpha I_{i,\alpha} \log k_\alpha \qquad \text{(S2)}$$

for the expected number of mentions allows for a capital status of a city is itself a result of optimization. We start with a more general assumption

$$\log m_{\mathrm{F},i} = \log c + a \log P_i \log P_i + bI_{i,cap} \log \frac{P_{i,c}}{P_i} + sI_{i,cap} + \sum_\alpha I_{i,\alpha} \log k_\alpha \qquad \text{(S3)}$$

implying that the capital status of a city might give either a constant (via parameter $s$) or population-dependent (via parameter $b$) boost to representation. It turned out that the second mechanism is enough to describe the observed data, i.e., assumption $s \neq 0$ does not pass the significance test. Furthermore, it turns out that $b \approx a/2$ and the assumption $b \neq a/2$ does not pass the significance test either.

*Outliers.* Partly due to the sparseness of the dataset, there is not a single city with $p < 0.0001$. There are 6 cities with $p < 0.001$, 5 of them are overmentioned, 1 is undermentioned, with clear individual reasons in all cases. The overmentioned cities are Accra (capital of the first decolonized Sub-Saharan African country and, as such, the focal point of the anticolonial movement in the late 1950s – early 1960s), Hiroshima (nuked in 1945), Santiago (attention related to the pro-Socialist activities of the Allende government and the subsequent anti-Allende coup), New York (location of the UN) and Stockholm (Sweden's traditional neutrality, as opposed to the USSR-guaranteed post-WWII neutrality of Finland and Austria, puts it into intermediate place between those two and the rest of Western Europe). Conversely, Madrid – the capital of a heavily anti-communist Franco regime – is strongly undermentioned.

*In-country city hierarchy.* In most cases the dataset is too sparse to probe the representation of the city hierarchy inside countries, except for the most over-represented ones. We summarize the data for non-capital cities of the 4 countries in the Capitalist I and Socialist I groups (Mongolia and Albania had no non-capital cities above 0.1 mln in 1970) in Table S8. It is notable that the model prediction of how mentions are split between the capital and other big cities seems to be consistently good.

*Berlin*. It is almost impossible to disentangle mentions of East and West Berlin. Indeed, (i) many mentions of Berlin in the dataset refer to the pre-World War II history, (ii) in many cases both sides of the divide are mentioned in a single story. For definiteness, we decided to use the population figure corresponding to the entirety of Berlin, and to treat it as capital of East Germany. We accept that this choice is imperfect but no better options seem available. However, readers should be aware that different choices will result in slight differences in the fitting results for East Germany.

*Albania.* Similarly, classification of Albania should be treated with caution: there is a single Albanian city (Tirana) in the dataset, and all its mentions happen before 1957, i.e., before Albania-Soviet split.

*Mongolia*. Similarly to Albania, there is a single Mongolian city (Ulaanbataar) in the dataset, unlike Tirana, the mentions of Ulaanbataar are evenly distributed through the dataset. Mongolia is notable as the only non-European country which is mentioned on par with the most mentioned European ones. It might be explained by a combination of the ideological conformity of the Mongolian regime, its close proximity to the Soviet Union and competition with China for the influence in Mongolia. However, given the sparseness of the dataset this is a relatively low-confidence result which needs further confirmation.

# Figures

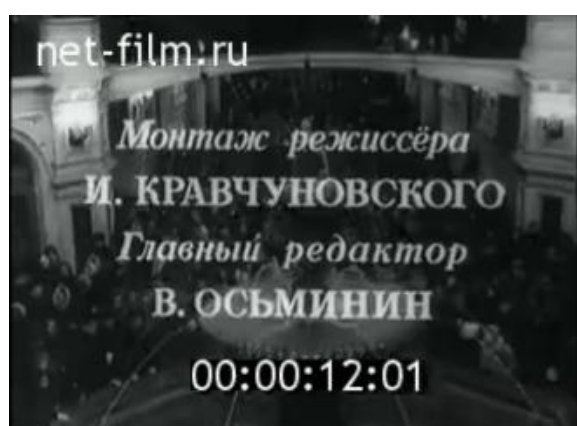

Novosti dnya 24/1954

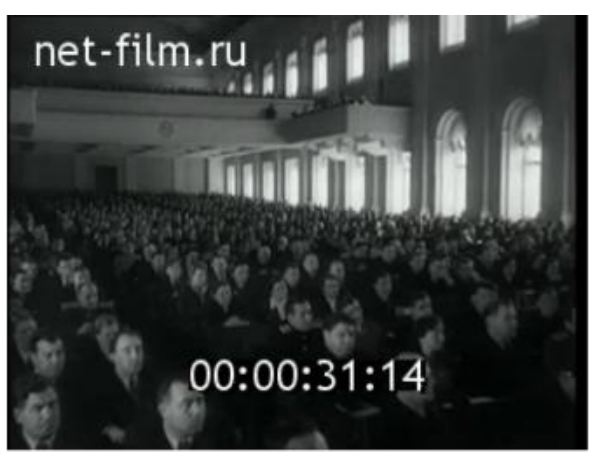

1. The first session of the Supreme Soviet of the USSR of the fourth convocation in the Kremlin.

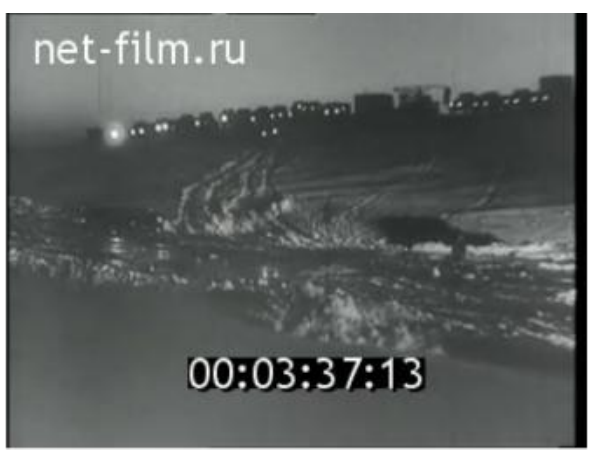

2. Tractor columns are going along the winter roads of Kazakhstan.

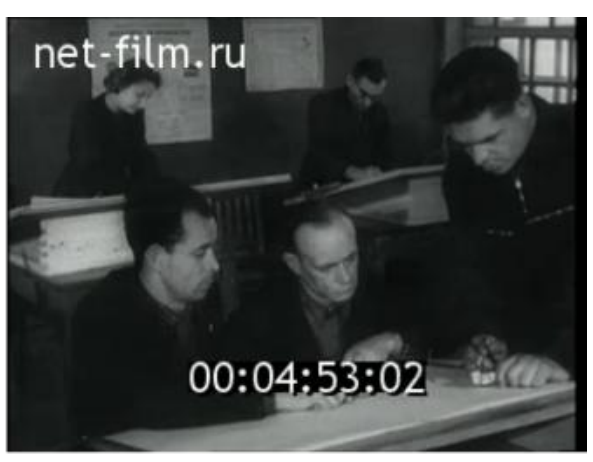

3. Boiler and fan plant in **Tula**. The inventor, turner Chekalin, works with a new cutter.

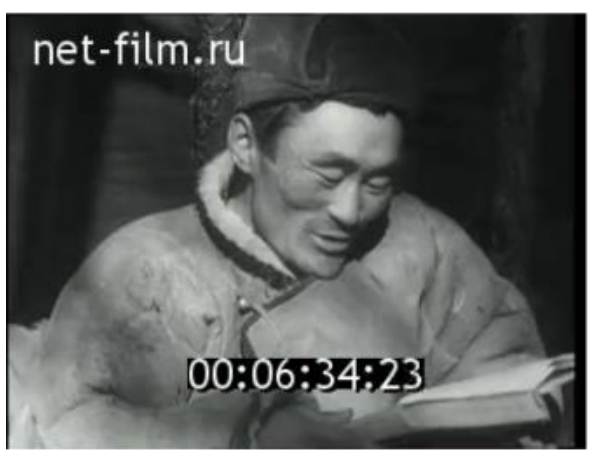

4. The librarian of a mountain village in the Sayan Mountains delivers books to tafalar reindeer breeders.

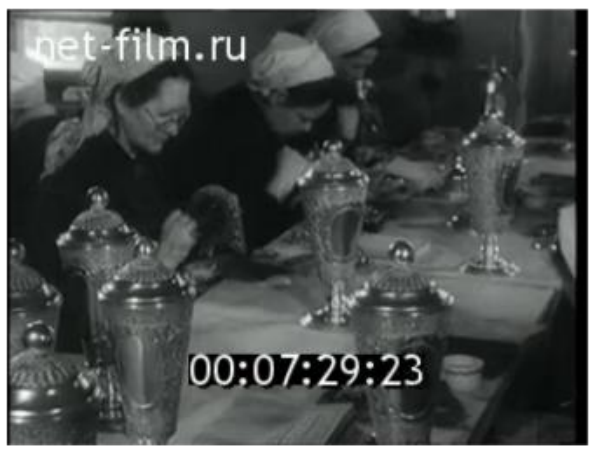

5. Jewelry trade in the village of Krasnoe, **Kostroma** region, jewelers at work.

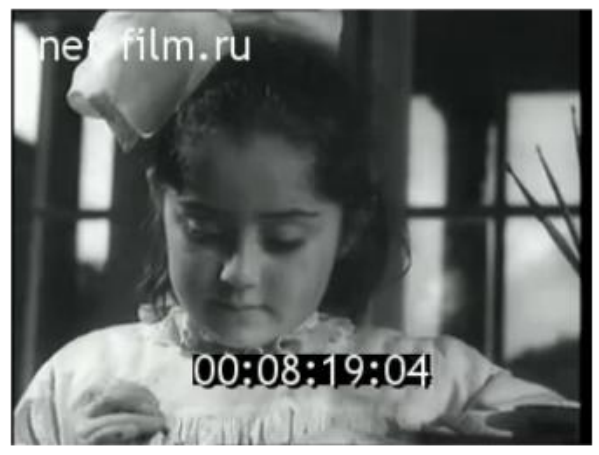

6. New kindergarten in **Tbilisi**.

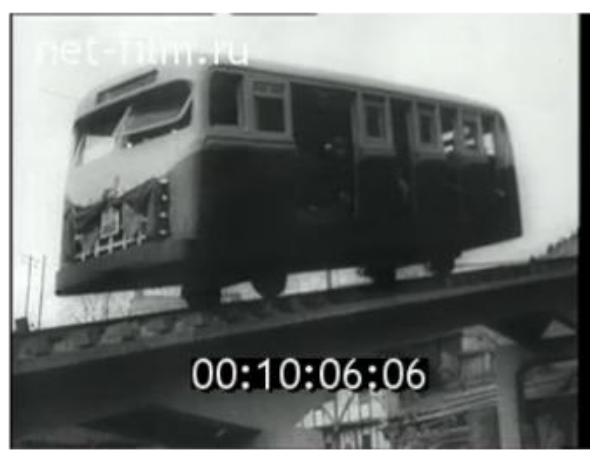

7. Construction of a funicular in the city of **Chongqing** in Southwest China, the townspeople travel in the funicular.

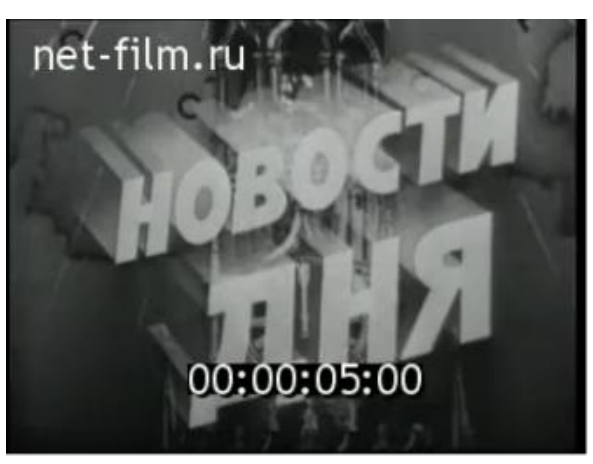

Novosti dnya 30 / 1970

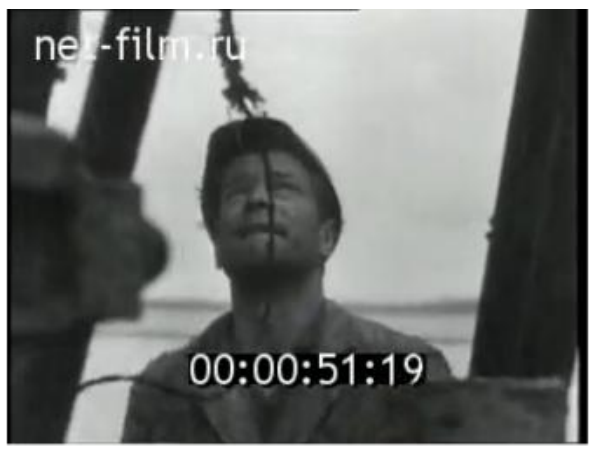

1. A view of oil rigs on Lake Samotlor in the **Tyumen** region. Oil workers are working at a well.

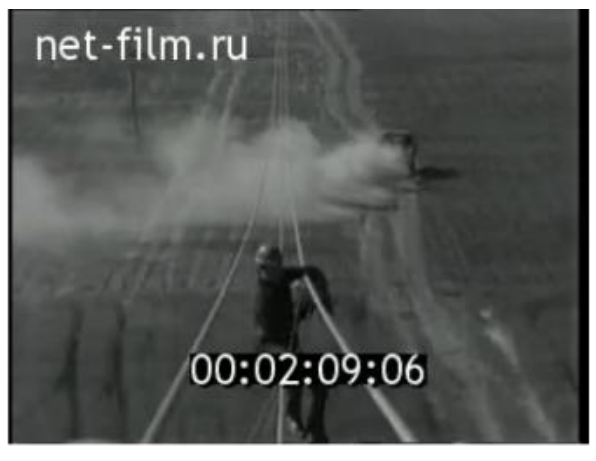

2. The electricians from Volgoelectrostroy in **Gorky** are raising the power line support.

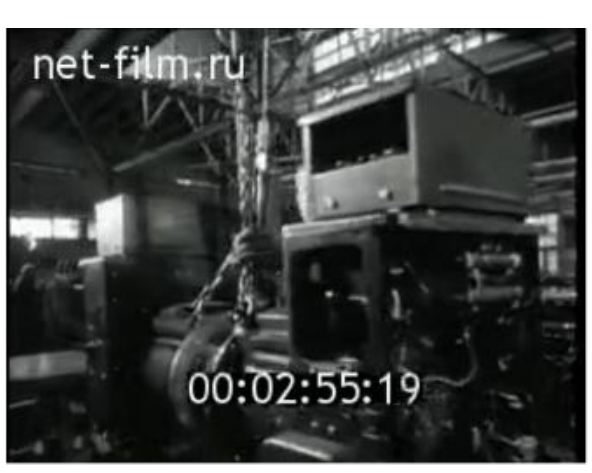

3. Production processes at the Vladimir Ilyich **Moscow** Electromechanical Plant.

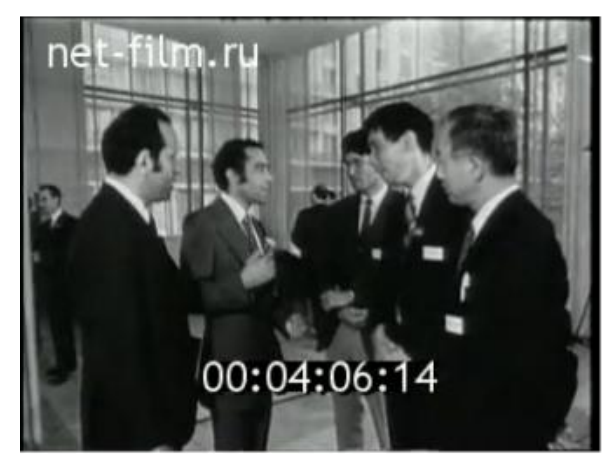

4. **Moscow** city. Speech by Deputy Chairman of the Council of Ministers of the USSR Z. N. Nuriev at the X International Congress of Soil Scientists at the Rossiya Hotel.

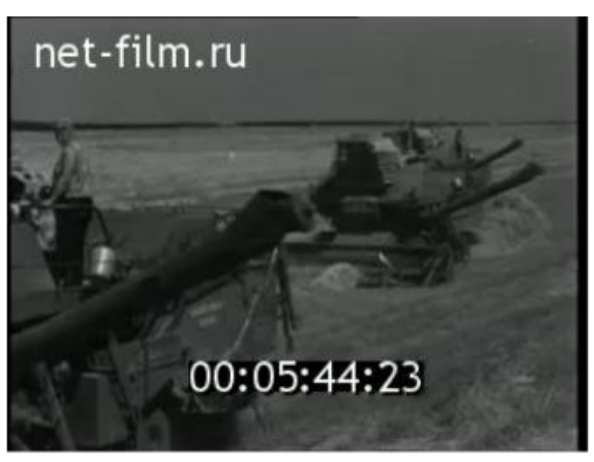

5. Agricultural processes in the collective farm "Lenin's Way" in the Peschanokopsky district of the **Rostov** region.

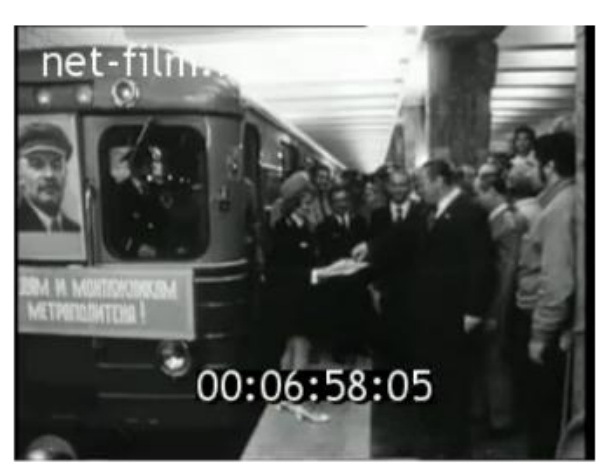

6. Builders work in a subway mine. A meeting of builders at the opening of the Belyaevo station of the **Moscow** Metro.

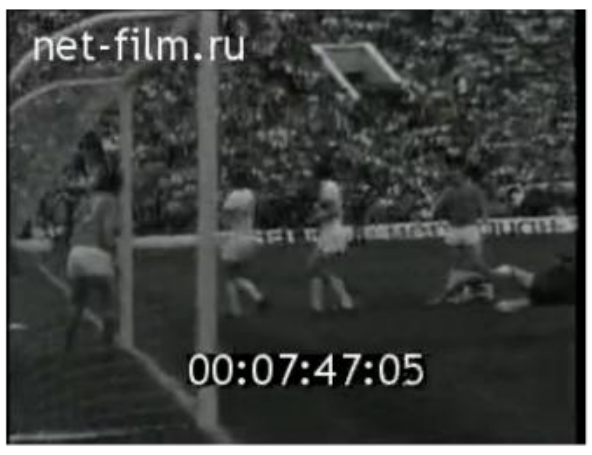

7. Moments of the final game of the USSR Cup on football between the teams "Dynamo" **Kiev** and "Zarya" **Voroshilovgrad**.

**Fig. S1.**

Snapshots from two exemplary newsreels, issue 24 of 1954 (top two rows) and issue 30 of 1970 (bottom two rows), with one snapshot per story. Snapshots are accompanied with Englis translations of the corresponding outlines, mentions of the cities are given in bold.

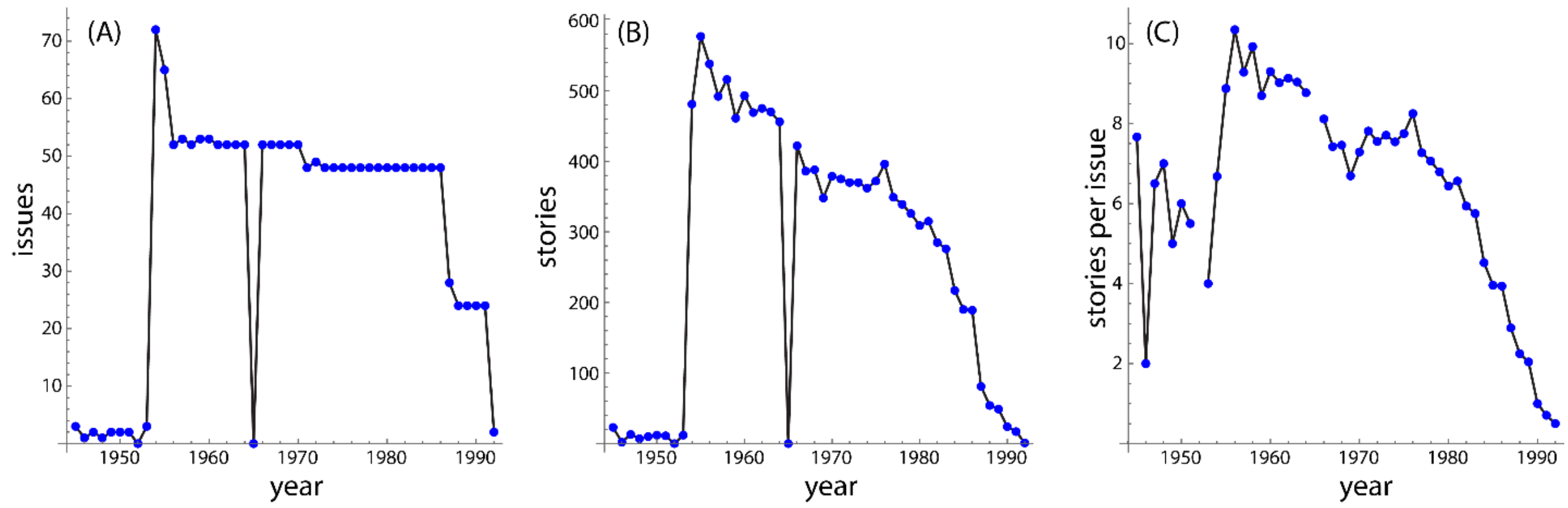

**Fig. S2.**

Temporal structure of the newsreel corpus used (A) number of issues in the dataset per year, (B) number of stories per year, (C) mean number of stories per issue.

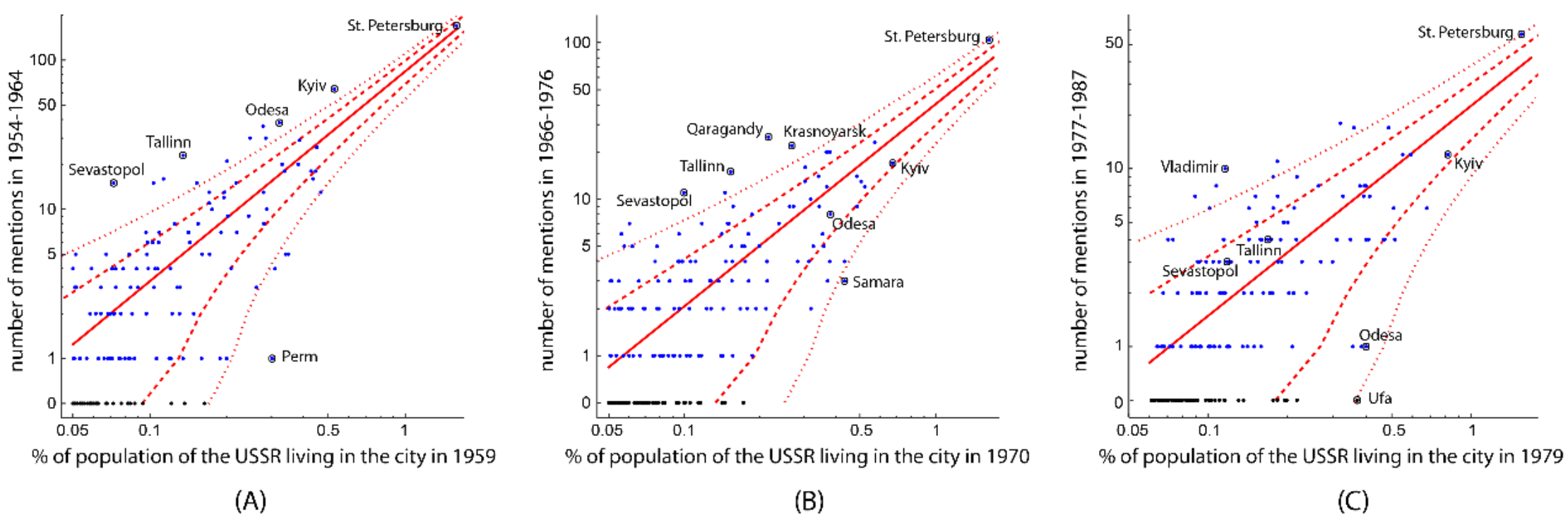

**Fig. S3.**

Scatter plots of the number of mentions vs population for the cities in the USSR for 3 periods of equal length: (A) mentions in 1954-64 vs population as of 1959 census; (B) mentions in 1966-76 vs population as of 1970 census; (C) mentions in 1977-87 vs population as of 1979 census. Red lines are best power-law fits with characteristics summarized in Table S3, dashed and dotted lines correspond to confidence intervals with $p = 0.05$ and $p = 0.001$, respectively. Cities with zero mentions (black dots) are shown out of scale. Selected cities are outlined, see discussion in the text.

*(Note for the referees: we provide this picture in higher resolution in a separate supplementary Figure.Time.pdf file).*

# Tables

| Main classification, done by MT, used for further analysis | Test alternative classification, done by KM and MO | |
|---|---|---|
| | Relevant (types 1, 2) | Irrelevant (types 3-5) |
| Relevant (types 1, 2) | 227 | 9 |
| Irrelevant (types 3-5) | 12 | 159 |

**Table S1.**

Results of the classification consistency test.

| Dataset | Number of cities | Cities with non-zero mentions | Number of mentions |
| --- | --- | --- | --- |
| Full | … | 180 | 879 |
| Above 1 mln | 135 | 62 | 598 |
| All cities of interest | 310 | 113 | 792 |

**Table S2.**

Mentions of cities of interest as compared to mentions of all cities outside the USSR.

| Cut-off | Number of cities | $a$ | $c$ |
|---|---|---|---|
| > 0.03% | 308 | $1.33 \pm 0.04$ | $1.34 \pm 0.13$ |
| > 0.03% + additionally at least 5 cities per republic | 328 | $1.33 \pm 0.04$ | $1.35 \pm 0.13$ |
| > 0.05% | 188 | $1.32 \pm 0.05$ | $1.38 \pm 0.15$ |
| > 0.1% | 81 | $1.37 \pm 0.07$ | $1.19 \pm 0.22$ |
| > 0.03% + Moscow | 309 | $1.82 \pm 0.03$ | $0.56 \pm 0.06$ |
| > 0.03% + Moscow with renormalized population | 309 | $1.175 \pm 0.015$ | $1.75 \pm 0.10$ |
| > 0.03% - St. Petersburg | 307 | $1.24 \pm 0.05$ | $1.53 \pm 0.13$ |

**Table S3.**

Parameters of the population-only model as function of the level of censoring cut-off. Influence of Moscow and St. Petersburg is also shown. *a* is the scaling exponent, *c* is the expected number of mentions for a city with 0.03% of population of the USSR

| Period | Cut-off | Number of cities | $a$ | $c$ |
|---|---|---|---|---|
| 1954-64 | 0.05% | 151 | $1.41 \pm 0.07$ | $0.61 \pm 0.10$ |
| 1966-76 | 0.05% | 194 | $1.29 \pm 0.08$ | $0.44 \pm 0.08$ |
| 1977-87 | 0.06% | 176 | $1.18 \pm 0.11$ | $0.36 \pm 0.08$ |

**Table S4.**

Parameters of the population-only model fitted separately for three periods of equal length: 1954-64, 1966-76 and 1977-87.

| Specialization | No of cities | Comments | Outcome |
|---|---|---|---|
| Capitals of Union level-republics | 14 | | Relevant, increases representation |
| Capitals of national autonomous republics inside Russia | 17 | | Relevant, decreases representation |
| Capitals of national autonomies outside Russia | 3 | | Irrelevant |
| Other cities located within national autonomies | 10 | | Irrelevant |
| Capitals of non-national regions (oblast or krai) | 112 | As of 1970 | Irrelevant |
| Port on the Black Sea | 12 | Location near the sea regardless of specialization (military, commerce, recreational, etc) | Relevant, increases representation, joined with Baltic and Pacific |
| Port on the Baltic Sea | 8 | See above | Relevant, increases representation, joined with Black Sea and Pacific |
| Port on the Pacific Ocean | 4 | See above | Relevant, increases representation, joined with Black Sea and Baltic |
| Port on the Azov Sea | 5 | See above | Irrelevant |
| Port on the Caspian Sea | 5 | See above | Irrelevant |
| Port on the Arctic coast | 4 | See above | Irrelevant |
| Resort city | 11 | Specialization in recreation regardless of seaside or inland location | Irrelevant |
| Hydroelectricity, big | 5 | Plants of >2 GW | Relevant, increases representation |
| Hydroelectricity, medium | 10 | Plants of 0.5-2 GW | Irrelevant |
| Steelworks | 18 | Full cycle only | Relevant, increases representation |
| Non-ferrous metallurgy | 18 | | Relevant, increases representation |
| Automobile plant | 14 | | Irrelevant |
| Coal mining | 38 | | Relevant, decreases representation |
| Newsreel-producing film studio | 26 | | Irrelevant |

**Table S5.**

Initial set of specializations used in the specialization model, with optimization outcomes for each of them. In order to avoid overfitting, only those 7 specializations, which are found to be relevant in the specialization model, are used in the full (specialization plus geolocation) model.

| City name | Contemporary city name, Cirillic | Specialization | Mentions | Mentions related to specialization | Fraction related to specialization |
|---|---|---|---|---|---|
| Odesa | Одесса | Seaside | 48 | 35 | 73% |
| Krasnoyarsk | Красноярск | Hydroelectricity | 47 | 25 | 53% |
| Tbilisi | Тбилиси | Republic capital | 38 | 10 | 26% |
| Cherepovets | Череповец | Steelwork | 18 | 18 | 100% |
| Kazan | Казань | Autonomous republic capital | 17 | 4 | 24% |
| Oskemen | Усть-Каменогорск | Non-ferrous metallurgy | 11 | 8 | 73% |
| Donetsk | Донецк/ Сталино | Coal | 12 | 4 | 33% |

**Table S6.**

Fractions of stories related to relevant city specialization for several selected cities.

| Model | Population | Geolocation | Specialization | Full |
| --- | --- | --- | --- | --- |
| No of cities | 308 | 328 | 328 | 328 |
| No of relevant parameters | 2 | 16 | 9 | 15 |
| $R^2$ | 0.905 | 0.952 | 0.945 | 0.964 |
| $\langle\sigma\rangle - 1$ | 3.08 | 1.89 | 2.12 | 1.53 |
| Cities with $p < 0.0001$ | 19 | 5 | 8 | 3 |
| Cities with $p < 0.0001$ | 30 | 14 | 13 | 7 |
| Cities with $p < 0.05$ | 69 | 58 | 52 | 41 |

**Table S7.**

Goodness of fit characteristics of the models for the cities in the USSR: coefficient of determination $R^2$, excess variation as compared to the Poisson distribution $\langle\sigma\rangle - 1$ (formula S1) and number of cities with significant deviations from the prediction of the model.

| Country | Cities | Mentions | Expected mentions |
|---|---|---|---|
| Austria | Vienna | 43 | 48.1 |
| | Graz, Linz, Salzburg, Innsbruck | 4 | 4.7 |
| Bulgaria | Sofia | 29 | 30.4 |
| | Plovdiv, Varna, Burgas, Ruse, Stara Zagora | 2 | 4.8 |
| Czechoslovakia | Prague | 51 | 47.2 |
| | Brno, Ostrava, Bratislava, Plzen, Kosice | 10 | 7.8 |
| Finland | Helsinki | 26 | 16.7 |
| | Turku, Tampere | 1 | 2.0 |

**Table S8.**

Mentions of non-capital cities in the four most over-represented countries compared to the model prediction and to the expectation for similar-sized cities in the Capitalist II ("over Europe") country group.